\documentclass[shortnames,nojss]{jss} % onecolumn

\usepackage[utf8]{inputenc}
\usepackage[T1]{fontenc}

\usepackage{booktabs}
\usepackage{multirow}

\usepackage{amsmath,amssymb,amsfonts}
\usepackage{bm}
\usepackage{dsfont}

\usepackage{pdflscape,rotating}
\usepackage{algorithm}

\shortcites{borghi_who,bmc_growth,schmid2013beta,villarini2009}

\DeclareSymbolFont{extraup}{U}{zavm}{m}{n}
\DeclareMathSymbol{\varheart}{\mathalpha}{extraup}{86}
\DeclareMathSymbol{\vardiamond}{\mathalpha}{extraup}{87}

\title{\pkg{gamboostLSS}: An \proglang{R} Package for Model Building and
  Variable Selection in the GAMLSS Framework}
\Plaintitle{gamboostLSS: An R Package for Model Building and Variable Selection
  in the GAMLSS Framework}
\Shorttitle{\pkg{gamboostLSS}: Model Building and Variable Selection for GAMLSS}
\author{Benjamin Hofner\\FAU Erlangen-Nürnberg \And
        Andreas Mayr\\FAU Erlangen-Nürnberg \And
        Matthias Schmid\\University of Bonn}
\Plainauthor{Benjamin Hofner, Andreas Mayr, Matthias Schmid}

\Abstract{ Generalized additive models for location, scale and shape (GAMLSS)
  are a flexible class of regression models that allow to model multiple
  parameters of a distribution function, such as the mean and the standard
  deviation, simultaneously. With the \proglang{R} package \pkg{gamboostLSS}, we
  provide a boosting method to fit these models. Variable selection and model
  choice are naturally available within this regularized regression framework.
  To introduce and illustrate the \proglang{R} package \pkg{gamboostLSS} and its
  infrastructure, we use a data set on stunted growth in India. In addition to
  the specification and application of the model itself, we present a variety of
  convenience functions, including methods for tuning parameter selection,
  prediction and visualization of results. The package \pkg{gamboostLSS} is
  available from CRAN (\url{http://cran.r-project.org/package=gamboostLSS}).}


\Address{
  Benjamin Hofner \& Andreas Mayr\\
  Department of Medical Informatics, Biometry and Epidemiology\\
  Friedrich-Alexander-Universität Erlangen-Nürnberg\\
  Waldstra{\ss}e 6\\
  91054 Erlangen, Germany\\
  E-mail: \email{benjamin.hofner@fau.de},\\
  \hphantom{E-mail: }\email{andreas.mayr@fau.de}\\
  URL: \url{http://www.imbe.med.uni-erlangen.de/cms/benjamin_hofner.html},\\
  \hphantom{URL: }\url{http://www.imbe.med.uni-erlangen.de/ma/A.Mayr/}\\

  Matthias Schmid\\
  Department of Medical Biometry, Informatics and Epidemiology\\
  University of Bonn\\
  Sigmund-Freud-Stra{\ss}e 25\\
  53105 Bonn\\
  E-mail: \email{schmid@imbie.meb.uni-bonn.de}\\
  URL: \url{https://www3.uni-bonn.de/imbie}
}

\begin{document}

\maketitle

\section{Introduction}
\label{sec:intro}

Generalized additive models for location, scale and shape (GAMLSS) are a
flexible statistical method to analyze the relationship between a response
variable and a set of predictor variables. Introduced by \citet{rs}, GAMLSS are
an extension of the classical GAM approach \citep{hastietib}. The main
difference between GAMs and GAMLSS is that GAMLSS do not only model the
conditional mean of the outcome distribution (location) but \textit{several} of
its parameters, including scale and shape parameters (hence the extension
``LSS''). In Gaussian regression, for example, the density of the outcome
variable $Y$ conditional on the predictors $\mathbf{X}$ may depend on the mean
parameter $\mu$, and an additional scale parameter $\sigma$, which corresponds
to the standard deviation of $Y| \mathbf{X}$. Instead of assuming $\sigma$ to be
fixed, as in classical GAMs, the Gaussian GAMLSS regresses both parameters on
the predictor variables,
\begin{eqnarray}
\label{Gaussian:mu}
\mu = \E(y \, | \,  \mathbf{X}) & = & \eta_\mu = \beta_{\mu,0} + \sum_j f_{\mu,j}(x_j), \\
\label{Gaussian:sigma}
\log(\sigma) = \log ( \sqrt{\VAR(y \, | \,  \mathbf{X})}) & = &
\eta_{\sigma} = \beta_{\sigma,0} +\sum_j f_{\sigma, j}(x_j),
\end{eqnarray}
where $\eta_\mu$ and $\eta_\sigma$ are \emph{additive predictors} with parameter
specific intercepts $\beta_{\mu,0}$ and $\beta_{\sigma,0}$, and functions
$f_{\mu,j}(x_j)$ and $f_{\sigma, j}(x_j)$, which represent the effect of
predictor $x_j$ on $\mu$ and $\sigma$, respectively. In this notation, the
functional terms $f(\cdot)$ can denote various types of effects (e.g., linear,
smooth, random).

In our case study, we will analyze the prediction of stunted growth for children
in India via a Gaussian GAMLSS. The response variable is a stunting score, which
is commonly used to relate the growth of a child to a reference population in
order to assess effects of malnutrition in early childhood. In our analysis, we
model the expected value ($\mu$) of this stunting score and also its variability
($\sigma$) via smooth effects for mother- or child-specific predictors, as well
as a spatial effect to account for the region of India, where the child is
growing up. This way, we are able to construct point predictors (via
$\eta_{\mu}$) and additionally child-specific prediction intervals (via
$\eta_{\mu}$ and $\eta_{\sigma}$) to evaluate the individual risk of stunted
growth.

In recent years, due to their versatile nature, GAMLSS have been used to address
research questions in a variety of fields. Applications involving GAMLSS range
from the normalization of complementary DNA microarray data
\citep{khondoker2009} and the analysis of flood frequencies
\citep{villarini2009} to the development of rainfall models
\citep{serinaldi2012} and stream-flow forecasting models \citep{ogtrop2011}. The
most prominent application of GAMLSS is the estimation of centile curves, e.g.,
for reference growth charts \citep{onis2006child, borghi_who, bmc_growth}. The
use of GAMLSS in this context has been recommended by the World Health
Organization \citep[see][and the references therein]{rigby:smoothing:2013}.
Classical estimation of a GAMLSS is based on backfitting-type Gauss-Newton
algorithms with AIC-based selection of relevant predictors. This strategy is
implemented in the \proglang{R} \citep{R:3.1.0} package \pkg{gamlss}
\citep{pkg:gamlss:4.3-0,gamlss:jss:2007}, which provides a great variety of
functions for estimation, hyper-parameter selection, variable selection and
hypothesis testing in the GAMLSS framework.

In this article we present the \proglang{R} package \pkg{gamboostLSS}
\citep{pkg:gamboostLSS:1.1-0}, which is designed as an alternative to
\pkg{gamlss} for high-dimensional data settings where variable selection is of
major importance. Specifically, \pkg{gamboostLSS} implements the
\emph{gamboostLSS} algorithm, which is a new fitting method for GAMLSS that was
recently introduced by \cite{mayretal}. The \emph{gamboostLSS} algorithm uses
the same optimization criterion as the Gauss-Newton type algorithms implemented
in the package \pkg{gamlss} (namely, the log-likelihood of the model under
consideration) and hence fits the same type of statistical model. In contrast to
\pkg{gamlss}, however, the \pkg{gamboostLSS} package operates within the
component-wise gradient boosting framework for model fitting and variable
selection \citep{BuehlmannYu2003,BuhlmannHothorn06}. As demonstrated in
\cite{mayretal}, replacing Gauss-Newton optimization by boosting techniques
leads to a considerable increase in flexibility: Apart from being able to fit
basically any type of GAMLSS, \pkg{gamboostLSS} implements an efficient
mechanism for variable selection and model choice. As a consequence,
\pkg{gamboostLSS} is a convenient alternative to the AIC-based variable
selection methods implemented in \pkg{gamlss}. The latter methods are known to
be unstable, especially when it comes to selecting possibly different sets of
variables for multiple distribution parameters. Furthermore, model fitting via
\emph{gamboostLSS} is also possible for high-dimensional data with more
candidate variables than observations ($p > n$), where the classical fitting
methods become unfeasible.

The \pkg{gamboostLSS} package is a comprehensive implementation of the most
important issues and aspects related to the use of the \emph{gamboostLSS}
algorithm. The package is available on CRAN
(\url{http://cran.r-project.org/package=gamboostLSS}). Current development
versions are hosted on R-forge
(\url{https://r-forge.r-project.org/projects/gamboostlss/}). As will be
demonstrated in this paper, the package provides a large number of response
distributions \citep[e.g., distributions for continuous data, count data and
survival data, including \textit{all} distributions currently available in the
\pkg{gamlss} framework; see][]{pkg:gamlss.dist:4.3-0}. Moreover, users of
\pkg{gamboostLSS} can choose among many different possibilities for modeling
predictor effects. These include linear effects, smooth effects and trees, as
well as spatial and random effects, and interaction terms.

After providing a brief theoretical overview of GAMLSS and component-wise
gradient boosting (Section~\ref{sec:boost-gamlss-models}), we will introduce the
\code{india} data set, which is shipped with the \proglang{R} package
\pkg{gamboostLSS} (Section~\ref{sec:india}). We present the infrastructure of
\pkg{gamboostLSS} and will show how the package can be used to build regression
models in the GAMLSS framework (Section~\ref{sec:package}). In particular, we
will give a step by step introduction to \pkg{gamboostLSS} by fitting a flexible
GAMLSS model to the \code{india} data. In addition, we will present a variety of
convenience functions, including methods for the selection of tuning parameters,
prediction and the visualization of results (Section~\ref{sec:methods}).

\section[Boosting GAMLSS models]{Boosting GAMLSS models}
\label{sec:boost-gamlss-models}

\emph{GamboostLSS} is an algorithm to fit GAMLSS models via
component-wise gradient boosting \citep{mayretal} adapting an earlier strategy
by \cite{schmidetal}. While the concept of boosting emerged from the field of
supervised machine learning, boosting algorithms are nowadays often applied as
flexible alternative to estimate and select predictor effects in statistical
models \citep[\textit{statistical boosting,}][]{mayr_boosting_part1}. The key
idea of statistical boosting is to iteratively fit the different predictors with
simple regression functions (base-learners) and combine the estimates to an
additive predictor. In case of gradient boosting, the base-learners are fitted
to the negative gradient of the loss function; this procedure can be described
as gradient descent in function space \citep{BuhlmannHothorn06}\footnote{For
  GAMLSS, we use the negative log-likelihood as loss function. Hence, the
  negative gradient of the loss functions equals the (positive) gradient of the
  log-likelihood. To avoid confusion we directly use the gradient of the
  log-likelihood in the remainder of the article.}.

To adapt the standard boosting algorithm to fit additive predictors for all
distribution parameters of a GAMLSS we extended the component-wise fitting to
multiple parameter dimensions: In each iteration, \emph{gamboostLSS} calculates
the partial derivatives of the log-likelihood function $l(y, \bm{\theta})$ with
respect to each of the additive predictors $\eta_{\theta_k}$, $k=1,\ldots,K$.
The predictors are related to the parameter vector $\bm{\theta} =
(\theta_k)^\top_{k = 1,\ldots,K}$ via parameter-specific link functions $g_k$,
$\theta_k = g_k^{-1}(\eta_{\theta_k})$. Typically, we have at maximum $K = 4$
distribution parameters \citep{rs}, but in principle more are possible. The
predictors are updated successively in each iteration. The current estimates of
the other distribution parameters are used as offset values. A schematic
representation of the updating process of \emph{gamboostLSS} with four
parameters in iteration $m+1$ looks as follows:

\begin{alignat*}{8}
   \frac{\partial}{\partial \eta_{\mu}}\, l(&y, \hat{\mu}^{[m]}&,& \hat{\sigma}^{[m]}&,& \hat{\nu}^{[m]}&,&
  \hat{\tau}^{[m]}&)
  &\quad \stackrel{\rm update}{\longrightarrow}&  \hat{\eta}_\mu^{[\boldmath{m+1}]}    \Longrightarrow  \hat{\mu}^{[\emph{m+1}]} \ ,& \\
   \frac{\partial}{\partial \eta_{\sigma}}\, l(&y, \hat{\mu}^{[\emph{m+1}]}&,& \hat{\sigma}^{[m]}&,& \hat{\nu}^{[m]}&,&
  \hat{\tau}^{[m]}&)
  &\quad \stackrel{\rm update }{\longrightarrow} \quad &  \hat{\eta}_\sigma^{[\emph{m+1}]}    \Longrightarrow  \hat{\sigma}^{[\emph{m+1}]} \ ,& \\
   \frac{\partial}{\partial \eta_{\nu}}\, l(&y, \hat{\mu}^{[\emph{m+1}]}&,& \hat{\sigma}^{[\emph{m+1}]}&,&
  \hat{\nu}^{[m]}&,& \hat{\tau}^{[m]}&)
  &\quad \stackrel{\rm update }{\longrightarrow}&  \hat{\eta}_\nu^{[\emph{m+1}]}    \Longrightarrow  \hat{\nu}^{[\emph{m+1}]} \ ,  \\
   \frac{\partial}{\partial \eta_{\tau}}\, l(&y, \hat{\mu}^{[\emph{m+1}]}&,& \hat{\sigma}^{[\emph{m+1}]}&,&
  \hat{\nu}^{[\emph{m+1}]}&,& \hat{\tau}^{[m]}&)
  &\quad \stackrel{\rm update}{\longrightarrow}& \hat{\eta}_\tau^{[\emph{m+1}]} \Longrightarrow
  \hat{\tau}^{[\emph{m+1}]} \ .
\end{alignat*}

The algorithm hence circles through the different parameter dimensions: in every
dimension, it carries out one boosting iteration, updates the corresponding
additive predictor and includes the new prediction in the loss function for the
next dimension.

As in classical statistical boosting, inside each boosting iteration only the
best fitting base-learner is included in the update. Typically, each
base-learner corresponds to one component of $\mathbf{X}$ and in every boosting
iteration only a small proportion (a typical value of the \textit{step-length}
is 0.1) of the fit of the selected base-learner is added to the current additive
predictor $\eta^{[m]}_{\theta_k}$. This procedure effectively leads to
data-driven variable selection which is controlled by the stopping iterations
$\bm{m}_{\text{stop}} = (m_{\text{stop},1},...,m_{\text{stop},K})^{\top}$: Each
additive predictor $\eta_{\theta_k}$ is updated until the corresponding stopping
iterations $\bm{m}_{\text{stop},k}$ is reached. If $m$ is greater than
$m_{\text{stop},k}$, the $k$th disribution parameter dimension is no longer
updated and simply skipped. Predictor variables that have never been selected up
to iteration $m_{\text{stop},k}$ are effectively excluded from the resulting
model. The vector $\bm{m}_{\text{stop}}$ is a tuning parameter that can, for
example, be determined using multi-dimensional cross-validation (see
Section~\ref{sec:model_tuning} for details). The complete \emph{gamboostLSS}
algorithm can be found in Appendix~\ref{algorithm} and is described in detail in
\cite{mayretal}.

\section{Childhood malnutrition in India}
\label{sec:india}

Eradicating extreme poverty and hunger is one of the Millennium Development
Goals that all 193 member states of the United Nations have agreed to achieve by
the year 2015. Yet, even in democratic, fast-growing emerging countries like
India, which is one of the biggest global economies, malnutrition of children is
still a severe problem in some parts of the population. Childhood malnutrition
in India, however, is not necessarily a consequence of extreme poverty but can
also be linked to low educational levels of parents and cultural factors
\citep{india_nut}. Following a bulletin of the WHO, growth assessment is the
best available way to define the health and nutritional status of children
\citep{who}. Stunted growth is defined as a reduced growth rate compared to a
standard population and is considered as the first consequence of malnutrition
of the mother during pregnancy, or malnutrition of the child during the first
months after birth. Stunted growth is often measured via a $Z$ score that
compares the anthropometric measures of the child with a reference population:
\begin{eqnarray*}
Z_i = \frac{\text{AI}_i - \text{MAI}}{s}
\end{eqnarray*}
In our case, the individual anthropometric indicator ($\text{AI}_i$) will be the
height of the child $i$, while MAI and $s$ are the median and the standard
deviation of the height of children in a reference population. This $Z$ score
will be denoted as \textit{stunting score} in the following. Negative values of
the score indicate that the child's growth is below the expected growth of a
child with normal nutrition.

The stunting score will be the outcome variable in our application,: we analyze
the relationship of the mother's and the child's BMI and age with stunted growth
resulting from malnutrition in early childhood. Furthermore, we will investigate
regional differences by including also the district of India in which the child
is growing up. The aim of the analysis is both, to explain the underlying
structure in the data as well as to develop a prediction model for children
growing up in India. A prediction rule, based also on regional differences,
could help to increase awareness for the individual risk of a child to suffer
from stunted growth due to malnutrition. For an in-depth analysis on the
multi-factorial nature of child stunting in India, based on boosted quantile
regression, see \citet{Fenske:2011:JASA}, and \citet{fenske2013plos}.

The data set that we use in this analysis is based on the Standard Demographic
and Health Survey, 1998-99, on malnutrition of children in India, which can be
downloaded after registration from \url{http://www.measuredhs.com}. For
illustrative purposes, we use a random subset of the original data set
containing 4000 observations (approximately 12\%) and only a
(very small) subset of variables. For details on the data set and the data
source see the manual of the \code{india} data set in the \pkg{gamboostLSS}
package and \citet{FahrmeirKneib:2011}.

\paragraph{Case study: Childhood malnutrition in India}

First of all we load the data sets \code{india} and \code{india.bnd} into the
workspace. The first data set includes the outcome and 5
explanatory variables. The latter data set consists of a special boundary file
containing the neighborhood structure of the districts in India.

\begin{Schunk}
\begin{Sinput}
R> library("gamboostLSS")
R> data("india")
R> data("india.bnd")
R> names(india)
\end{Sinput}
\begin{Soutput}
[1] "stunting"   "cbmi"       "cage"       "mbmi"       "mage"      
[6] "mcdist"     "mcdist_lab"
\end{Soutput}
\end{Schunk}

\begin{figure}[ht!]
  \centering
\includegraphics{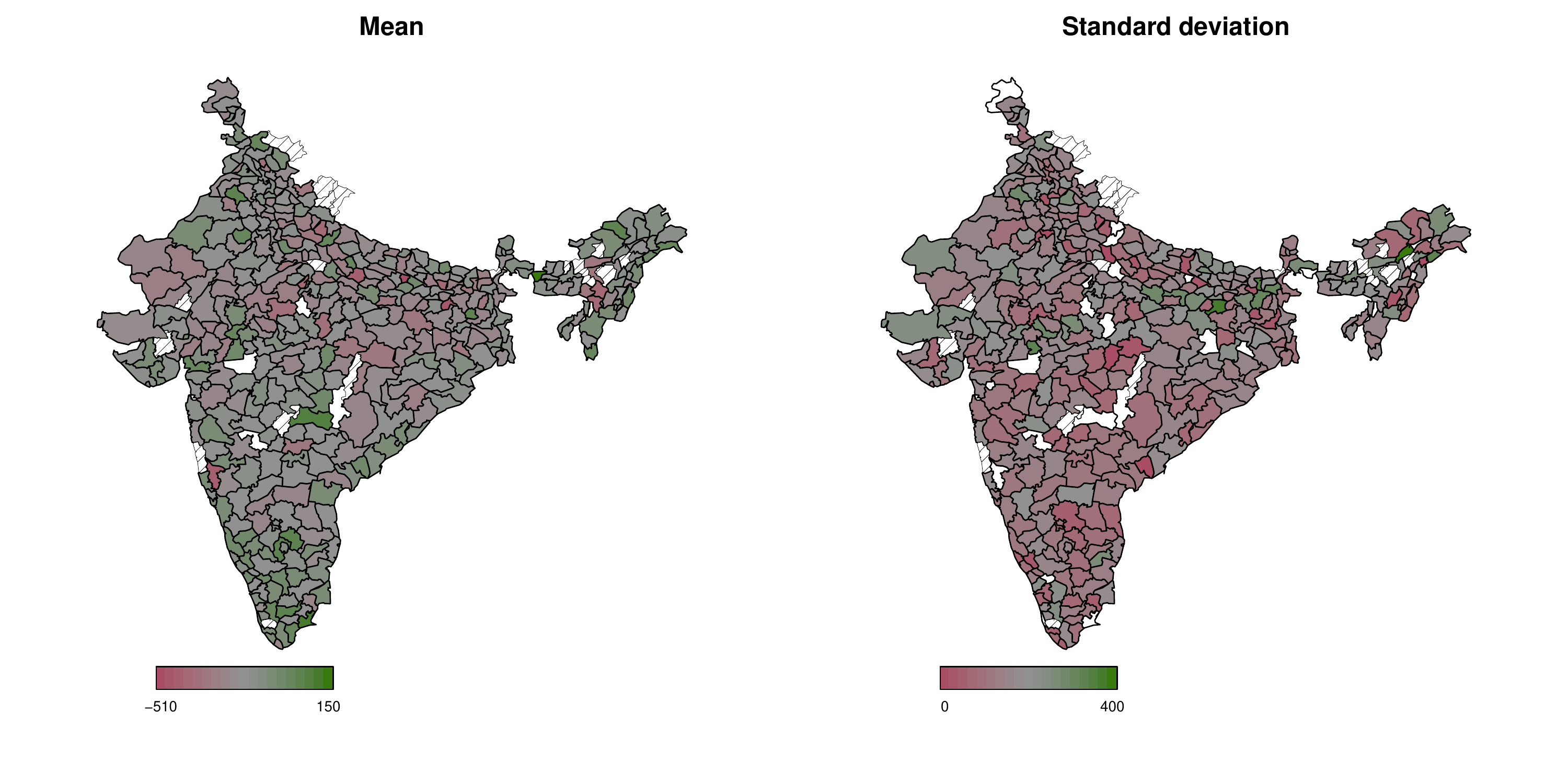}
\caption{Spatial structure of stunting in India. The raw mean per district is
  given in the left figure, ranging from dark red (low stunting score), to dark
  green (higher scores). The right figure depicts the standard deviation of the
  stunting score in the district, ranging from dark red (no variation) to dark
  green (maximal variability). Dashed regions represent regions without
  data.}\label{fig:india}
\end{figure}

The outcome variable \code{stunting} is depicted with its spatial structure in
Figure~\ref{fig:india}. An overview of the data set can be found in
Table~\ref{tab:summary_india}. One can clearly see a trend towards malnutrition
in the data set as even the 75\% quantile of the stunting score is below zero.
\hfill $\vardiamond$

\setkeys{Gin}{width = 0.90\textwidth}

\begin{table}[h!]
  \centering
  \caption{Overview of \code{india} data.\label{tab:summary_india}}
  \begin{tabular}{llrrrrrr}
\toprule  &  & Min. & 25\% Qu. & Median & Mean & 75\% Qu. & Max. \\ 
 \cmidrule{3-8} Stunting & \code{stunting} & -599.00 & -287.00 & -176.00 & -175.41 & -65.00 & 564.00  \\ 
BMI (child) & \code{cbmi} &   10.03 &   14.23 &   15.36 &   15.52 &  16.60 &  25.95  \\ 
Age (child; months) & \code{cage} &    0.00 &    8.00 &   17.00 &   17.23 &  26.00 &  35.00  \\ 
BMI (mother) & \code{mbmi} &   13.14 &   17.85 &   19.36 &   19.81 &  21.21 &  39.81  \\ 
Age (mother; years) & \code{mage} &   13.00 &   21.00 &   24.00 &   24.41 &  27.00 &  49.00  \\ \bottomrule   \end{tabular}
\end{table}
\pagebreak[3]

\section[The Package gamboostLSS]{The package \pkg{gamboostLSS}}
\label{sec:package}
The \emph{gamboostLSS} algorithm is implemented in the publicly available
\proglang{R} add-on package \pkg{gamboostLSS} \citep{pkg:gamboostLSS:1.1-0}. The
package makes use of the fitting algorithms and some of the infrastructure of
\pkg{mboost} \citep{pkg:mboost:2.3-0}. Furthermore, many naming conventions and
features are implemented in analogy to \pkg{mboost}. By relying on the
\pkg{mboost} package, \pkg{gamboostLSS} incorporates a wide range of
base-learners and hence offers a great flexibility when it comes to the types of
predictor effects on the parameters of a GAMLSS distribution. In addition to
making the infrastructure available for GAMLSS, \pkg{mboost} constitutes a
well-tested, mature software package in the back end. For the users of
\pkg{mboost}, \pkg{gamboostLSS} offers the advantage of providing a drastically
increased number of possible distributions to be fitted by boosting.

As a consequence of this partial dependency on \pkg{mboost}, we recommend users
of \pkg{gamboostLSS} to make themselves familiar with the former before using
the latter package. To make this tutorial self-contained, we try to shortly
explain all relevant features here as well. However, a dedicated hands-on
tutorial is available for an applied introduction to \pkg{mboost}
\citep{Hofner:mboost:2014}.

\subsection{Model-fitting}\label{sec:model-fitting}

The models can be fitted using the function \code{glmboostLSS()} for linear
models. For all kinds of structured additive models the function
\code{gamboostLSS()} can be used. The function calls are as
follows\footnote{Note that here and in the following we sometimes restrict the
  focus to the most important or most interesting arguments of a function.
  Further arguments might exist. Thus, for a complete list of arguments and
  their description we refer the reader to the respective manual.}:

\begin{Sinput}
    glmboostLSS(formula, data = list(), families = GaussianLSS(),
                control = boost_control(), weights = NULL, ...)
    gamboostLSS(formula, data = list(), families = GaussianLSS(),
                control = boost_control(), weights = NULL, ...)
\end{Sinput}

The \code{formula} can consist of a single \code{formula} object, yielding the
same candidate model for all distribution parameters. For example,
\begin{Sinput}
R> glmboostLSS(y ~ x1 + x2 + x3 + x4, data = data)
\end{Sinput}
specifies linear models with predictors \code{x1} to \code{x4} for all GAMLSS
parameters (here $\mu$ and $\sigma$ of the Gaussian distribution). As an
alternative, one can also use a named list to specify different candidate models
for different parameters, e.g.
\begin{Sinput}
R> glmboostLSS(list(mu = y ~ x1 + x2, sigma = y ~ x3 + x4), data = data)
\end{Sinput}
fits a linear model with predictors \code{x1} and \code{x2} for the \code{mu}
component and a linear model with predictors \code{x3} and \code{x4} for the
\code{sigma} component. As for all \proglang{R} functions with a formula
interface, one must specify the data set to be used (argument \code{data}).
Additionally, \code{weights} can be specified for weighted regression. Instead
of specifying the argument \code{family} as in \pkg{mboost} and other modeling
packages, the user needs to specify the argument \code{families}, which
basically consists of a list of sub-families, i.e., one family for each of the
GAMLSS distribution parameters. These sub-families define the parameters of the
GAMLSS distribution to be fitted. Details are given in the next section.

The initial number of boosting iterations as well as the step-lengths
($\nu_{\text{sl}}$; see Appendix~\ref{algorithm}) are specified via the function
\code{boost\_control()} with the same arguments as in \pkg{mboost}. However, in
order to give the user the possibility to choose different values for each
additive predictor (corresponding to the different parameters of a GAMLSS), they
can be specified via a vector or list\footnote{Preferably a named vector or list
  should be used where the names correspond to the names of the sub-families.}.
For example, one can specify:

\begin{Sinput}
R> boost_control(mstop = c(mu = 100, sigma = 200),
R>               nu = c(mu = 0.2, sigma = 0.01))
\end{Sinput}

Specifying a single value for the stopping iteration \code{mstop} or the
step-length \code{nu} results in equal values for all sub-families. The defaults
is \code{mstop = 100} for the initial number of boosting iterations and \code{nu
  = 0.1} for the step-length. Additionally, the user can specify if status
information should be printed by setting \code{trace = TRUE} in
\code{boost_control}.

\subsection{Distributions}\label{sec:distributions}

Some GAMLSS distributions are directly implemented in the \proglang{R} add-on
package \pkg{gamboostLSS} and can be specified via the \code{families} argument
in the fitting function \code{gamboostLSS()} and \code{glmboostLSS()}. An
overview of the implemented families is given in
Table~\ref{tab:gamboostlss_families}. The parametrization of the negative
binomial distribution, the log-logistic distribution and the $t$ distribution in
boosted GAMLSS models is given in \citet{mayretal}. The derivation of
boosted beta regression, another special case of GAMLSS, can for example be
found in \cite{schmid2013beta}. In our case study we will use the default
\code{GaussianLSS()} family to model childhood malnutrition in India. The
resulting object of the family looks as follows:

\begin{Schunk}
\begin{Sinput}
R> str(GaussianLSS(), 1)
\end{Sinput}
\begin{Soutput}
List of 2
 $ mu   :Formal class 'boost_family' [package "mboost"] with 10 slots
\end{Soutput}
\end{Schunk}

\begin{landscape}
  \begin{table}[h!]
    \centering
    \caption{Overview of \code{"families"} that are implemented in
      \pkg{gamboostLSS}. For every distribution parameter the  corresponding
      link-function is displayed (id = identity link).} \label{tab:gamboostlss_families}
    \begin{tabular}{lllllllp{0.5\textwidth}}
      \toprule
      & &  Name & Response & $\mu$ & $\sigma$  & $\nu$  & Note\\
      \cmidrule{2-8}

      \multicolumn{8}{l}{\textbf{Continuous response}} \\
      & Gaussian & \code{GaussianLSS()} & cont.  & id & log & & \\
      & Student's $t$  & \code{StudentTLSS()} & cont.   & id & log & log & The 3rd parameter is denoted by \code{df} (degrees of freedom). \\
      \cmidrule{2-8}

      \multicolumn{8}{l}{\textbf{Continuous non-negative response}} \\
      & Gamma & \code{GammaLSS()} & cont. $>0$  & log & log & & \\
      \cmidrule{2-8}

      \multicolumn{8}{l}{\textbf{Fractions and bounded continuous response}} \\
      & Beta & \code{BetaLSS()} & $\in (0,1)$  & logit & log & & The 2nd parameter is denoted by \code{phi}.\\
      \cmidrule{2-8}

      \multicolumn{8}{l}{\textbf{Models for count data}} \\
      & Negative binomial & \code{NBinomialLSS()} & count  & log & log &   & For over-dispersed count data.\\
      & Zero inflated Poisson & \code{ZIPoLSS()} & count  & log & logit & & For zero-inflated count data; the 2nd parameter is the probability parameter of the zero mixture component.\\
      & Zero inflated neg. binomial & \code{ZINBLSS()} & count  & log & log & logit & For over-dispersed and zero-inflated count data; the 3rd parameter is the probability parameter of the zero mixture component.\\
      \cmidrule{2-8}

      \multicolumn{8}{l}{\textbf{Survival models} \citep[accelerated failure time models; see, e.g.,][]{klein03}} \\
      & Log-normal & \code{LogNormalLSS()} & cont. $>0$ & id & log &&
      \multirow{3}{8cm}{All three families assume that the data are subject to
        right-censoring. Therefore the response must be a \code{Surv()} object.}\\
      & Weibull & \code{WeibullLSS()} & cont. $>0$ & id & log &&  \\
      & Log-logistic & \code{LogLogLSS()} & cont. $>0$ & id & log &&\\
      \bottomrule
  \end{tabular}
\end{table}
\end{landscape}

\begin{Schunk}
\begin{Soutput}
 $ sigma:Formal class 'boost_family' [package "mboost"] with 10 slots
 - attr(*, "class")= chr "families"
 - attr(*, "qfun")=function (p, mu = 0, sigma = 1, lower.tail = TRUE, log.p = FALSE)  
 - attr(*, "name")= chr "Gaussian"
\end{Soutput}
\end{Schunk}

We obtain a list of class \code{"families"} with two sub-families, one for the
$\mu$ parameter of the distribution and one for the $\sigma$ parameter. Each of
the sub-families is of type \code{"boost_family"} from package \pkg{mboost}.
Attributes specify the name and the quantile function (\code{"qfun"}) of the
distribution.

In addition to the families implemented in the \pkg{gamboostLSS} package, there
are many more possible GAMLSS distributions available in the \pkg{gamlss.dist}
package \citep{pkg:gamlss.dist:4.3-0}. In order to make our boosting approach
available for these distributions as well, we provide an interface to
automatically convert available distributions of \pkg{gamlss.dist} to objects of
class \code{"families"} to be usable in the boosting framework via the function
\code{as.families()}. As input, a character string naming the
\code{"gamlss.family"}, or the function itself is required. The function
\code{as.families()} then automatically constructs a \code{"families"} object
for the \pkg{gamboostLSS} package. To use for example the gamma family as
parametrized in \pkg{gamlss.dist}, one can simply use \code{as.families("GA")}
and plug this into the fitting algorithms of \pkg{gamboostLSS}:

\begin{Sinput}
R> gamboostLSS(y ~ x, families = as.families("GA"))
\end{Sinput}

With this interface, it is possible to apply boosting for any distribution
implemented in \pkg{gamlss.dist} and for all new distributions that will be
added in the future. Note that one can also fit censored or truncated
distributions by using \code{gen.cens()} (from package \pkg{gamlss.cens}) or
\code{gen.trun()} (from package \pkg{gamlss.tr}), respectively. An overview of
common GAMLSS distributions is given in Appendix~\ref{sec:additional-families}.
Minor differences in the model fit when applying a pre-specified distribution
(e.g., \code{GaussianLSS()}) and the transformation of the corresponding
distribution from \pkg{gamlss.dist} (e.g., \code{as.families("NO")}) can be
explained by possibly different offset values.

\subsection{Base-learners}\label{sec:base-learners}

For the base-learners, which carry out the fitting of the gradient vectors using
the covariates, the \pkg{gamboostLSS} package totally depends on the
infrastructure of \pkg{mboost}. Hence, every base-learner which is available in
\pkg{mboost} can also be applied to fit GAMLSS distributions via
\pkg{gamboostLSS}. The choice of base-learners is crucial for the application of
the \emph{gamboostLSS} algorithm, as they define the type(s) of effect(s) that
covariates will have on the predictors of the GAMLSS distribution parameters.

The available base-learners of \pkg{mboost}\footnote{See
  \citet{Hofner:mboost:2014} for details and application notes.} include simple
linear models for \textit{linear} effects and penalized regression splines
\citep[\textit{P}-splines, ][]{Eilers1996} for \textit{non-linear} effects.
\textit{Spatial} or other \textit{bivariate} effects can be incorporated by
setting up a bivariate tensor product extension of P-splines for two continuous
variables \citep{kneibetal}. Another way to include spatial effects is the
adaptation of Markov random fields for modeling a neighborhood structure
\citep{sobotka12} or radial basis functions \citep{Hofner:Dissertation:2011}.
\textit{Constrained} effects such as monotonic or cyclic effects can be
specified as well \citep{Hofner:monotonic:2011,Hofner:constrained:2014}.
\textit{Random} effects can be taken into account by using ridge-penalized
base-learners for fitting categorical grouping variables such as random
intercepts or slopes \citep[see supplementary material of][]{kneibetal}.

\paragraph{Case study (ctd.): Childhood malnutrition in India}

Before fitting a model for stunting in India, we rescale the outcome variable
\code{stunting} to increase the convergence speed of the algorithm. For details
and recommendations see Appendix~\ref{sec:stab-ngrad}.

\begin{Schunk}
\begin{Sinput}
R> india$stunting_rs <- india$stunting/600
\end{Sinput}
\end{Schunk}

In the next step, we are going to set up and fit our model. Usually, one could
use \code{bmrf(mcdist, bnd = india.bnd)} to specify the spatial base-learner
using a Markov random field. However, as it is relatively time-consuming to
compute the neighborhood matrix from the boundary file and as we need it several
times, we pre-compute it once. Note that \pkg{BayesX} \citep{pkg:BayesX:0.2-8}
needs to be loaded in order to use this function:

\begin{Schunk}
\begin{Sinput}
R> library("BayesX")
R> neighborhood <- bnd2gra(india.bnd)
\end{Sinput}
\end{Schunk}

The other effects can be directly specified without special care. We use smooth
effects for the age (\code{mage}) and BMI (\code{mbmi}) of the mother and smooth
effects for the age (\code{cage}) and BMI (\code{cbmi}) of the child. Finally,
we specify the spatial effect for the district in India where mother and child
live (\code{mcdist}).

\begin{Schunk}
\begin{Sinput}
R> ctrl <- boost_control(trace = TRUE, mstop = c(mu = 193, sigma = 41))
R> mod <- gamboostLSS(stunting_rs ~ bbs(mage) + bbs(mbmi) +
+                       bbs(cage) + bbs(cbmi) +
+                       bmrf(mcdist, bnd = neighborhood),
+                     data = india,
+                     families = GaussianLSS(),
+                     control = ctrl)
\end{Sinput}
\begin{Soutput}
[   1] ................................. -- risk: -76.91 
[  36] ................................. -- risk: -118.0861 
[  71] ................................. -- risk: -144.3796 
[ 106] ................................. -- risk: -164.923 
[ 141] ................................. -- risk: -181.4993 
[ 176] ................
Final risk: -188.6989 
\end{Soutput}
\end{Schunk}

We specified the initial number of boosting iterations as \code{mstop = c(mu =
  193, sigma = 41)}, i.e., we used $193$ boosting iterations for the $\mu$
parameter and only $41$ for the $\sigma$ parameter. This means that we cycle
between the $\mu$ and $\sigma$ parameter until we have computed $41$ updates
steps in both sub-models. Subsequently, we update only the $\mu$ model and leave
the $\sigma$ model unchanged. The selection of these tuning parameters will be
discussed in the next Section. \hfill $\vardiamond$\\

Instead of optimizing the gradients per GAMLSS-parameter in each
boosting iteration, one can potentially stabilize the estimation further by
standardizing the gradients in each step. Details and an explanation
are given in Appendix~\ref{sec:stab-ngrad}.

\paragraph{Case study (ctd.): Childhood malnutrition in India}

We now use the built-in stabilization and refit the model:

\begin{Schunk}
\begin{Sinput}
R> options(gamboostLSS_stab_ngrad = TRUE)
R> mod <- gamboostLSS(stunting_rs ~ bbs(mage) + bbs(mbmi) +
+                       bbs(cage) + bbs(cbmi) +
+                       bmrf(mcdist, bnd = neighborhood),
+                     data = india,
+                     families = GaussianLSS(),
+                     control = ctrl)
\end{Sinput}
\begin{Soutput}
[   1] ................................. -- risk: -64.01392 
[  36] ................................. -- risk: -95.13413 
[  71] ................................. -- risk: -112.5789 
[ 106] ................................. -- risk: -126.7645 
[ 141] ................................. -- risk: -139.0128 
[ 176] ................
Final risk: -144.7426 
\end{Soutput}
\end{Schunk}

One can clearly see that the stabilization changes the model if we look at the
intermediate and final risks. \hfill $\vardiamond$

\subsection{Model tuning: Early stopping to prevent overfitting}\label{sec:model_tuning}

As for other component-wise boosting algorithms, the most important tuning
parameter of the \emph{gamboostLSS} algorithm is the stopping iteration
$\bm{m}_{\text{stop}}$ (here a $K$-dimensional vector). In some low-dimensional
settings it might be convenient to let the algorithm run until convergence
(i.e., use a large number of iterations for each of the $K$ distribution
parameters). In this case, as they are optimizing the same likelihood, boosting
should converge to the same model as \pkg{gamlss} -- at least when the same
penalties are used for smooth effects.

However, in most settings, where the application of boosting is favorable, it
will be crucial that the algorithm is not run until convergence but some sort of
early stopping is applied \citep{Mayr:mstop:2012}. Early stopping results in
shrunken effect estimates, which has the advantage that predictions become more
stable since the variance of the estimates is reduced. Another advantage of
early stopping is that \emph{gamboostLSS} has an intrinsic mechanism for
data-driven variable selection, since only the best-fitting base-learner is
updated in each boosting iteration. Hence, the stopping iteration
$m_{\text{stop},k}$ does not only control the amount of shrinkage applied to the
effect estimates but also the complexity of the models for the distribution
parameter $\theta_k$.

To find the optimal complexity, the resulting model should be evaluated
regarding the predictive risk on a large grid of stopping values by
cross-validation or resampling methods, using the function \code{cvrisk()}. In
case of \emph{gamboostLSS}, the predictive risk is computed as the negative log
likelihood of the out-of-bag sample. The search for the optimal
$\bm{m}_{\text{stop}}$ based on resampling is far more complex than for standard
boosting algorithms. Different stopping iterations can be chosen for the
parameters, thus allowing for different levels of complexity in each sub-model
(\emph{multi-dimensional} early stopping). In the package \pkg{gamboostLSS} a
multi-dimensional grid can be easily created utilizing the function
\code{make.grid()}.

As already stated by the inventors of GAMLSS, in most of the cases the
$\mu$ parameter is of greatest interest in a GAMLSS model and thus more care
should be taken to accurately model this parameter\footnote{``Respect the parameter
  hierarchy when you are fitting a model. For example a good model for $\mu$
  should be fitted before a model for $\sigma$ is fitted.'' \citep[help page for
  the function \code{gamlss()},][]{pkg:gamlss:4.3-0}}. Consequently, we provide an
option \code{dense_mu_grid} in the \code{make.grid()} function that allows to have a
finer grid for (a subset of) the $\mu$ parameter. Thus, we can better tune the
complexity of the model for $\mu$ which helps to avoid over- or underfitting of
the mean without relying to much on the grid. Details and explanations are given
in the following paragraphs.

\paragraph{Case study (ctd.): Childhood malnutrition in India}

We first set up a grid for \code{mstop} values starting at $20$ and going in
$10$ equidistant steps on a logarithmic scale to $500$.

\begin{Schunk}
\begin{Sinput}
R> grid <- make.grid(max = c(mu = 500, sigma = 500), min = 20,
+                    length.out = 10, dense_mu_grid = FALSE)
\end{Sinput}
\end{Schunk}

Additionally, we can use the \code{dense\_mu\_grid} option to create a dense
grid for $\mu$. This means that we compute the risk for all iterations
$m_{\text{stop},\mu}$, if $m_{\text{stop},\mu} \geq m_{\text{stop},\sigma}$ and
do not use the values on the sparse grid only:

\begin{Schunk}
\begin{Sinput}
R> densegrid <- make.grid(max = c(mu = 500, sigma = 500), min = 20,
+                         length.out = 10, dense_mu_grid = TRUE)
R> plot(densegrid, pch = 20, cex = 0.2)
R> abline(0,1)
R> points(grid, pch = 20, col = "red")
\end{Sinput}
\end{Schunk}

\begin{figure}[h!]
  \centering
\includegraphics{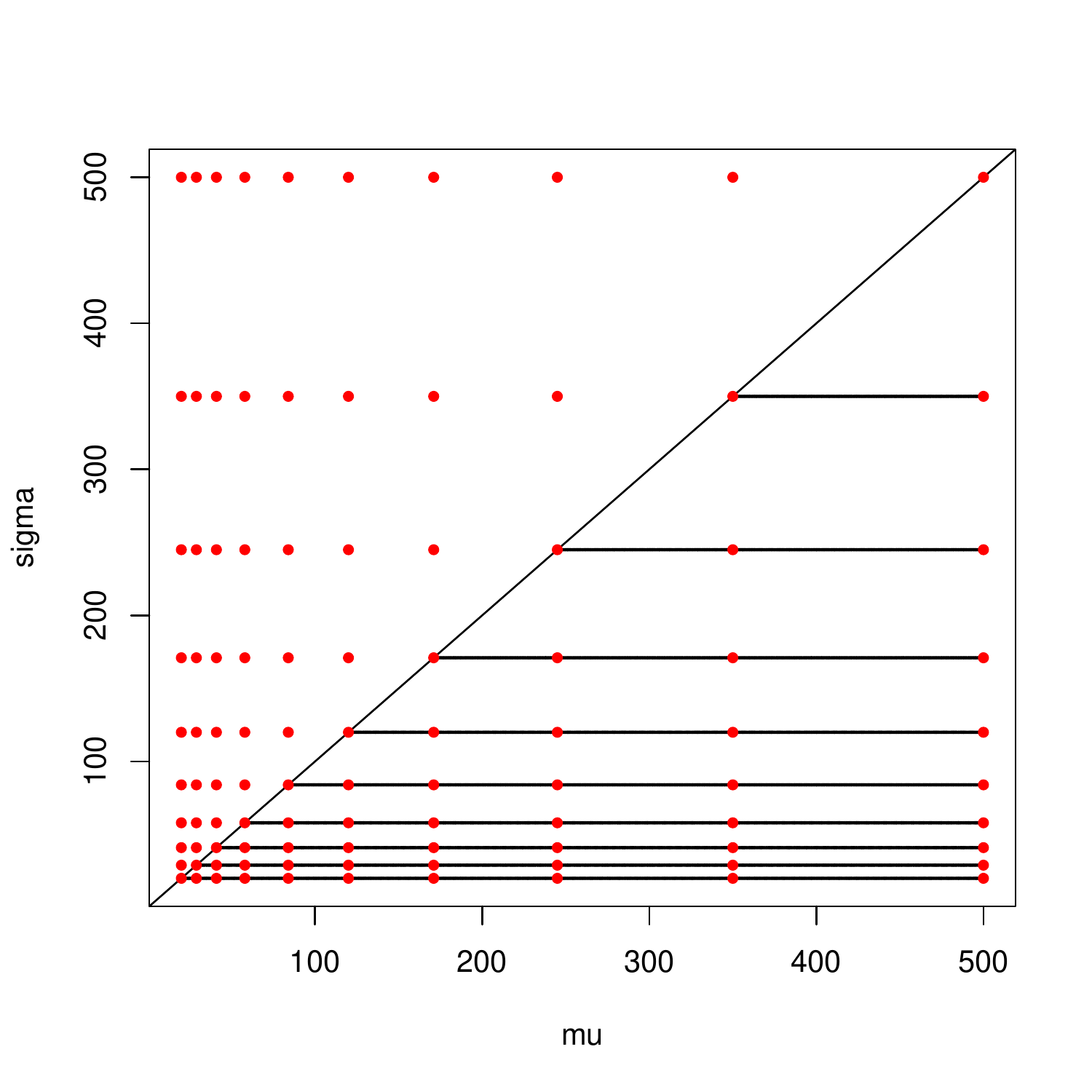}
\hfill
\includegraphics{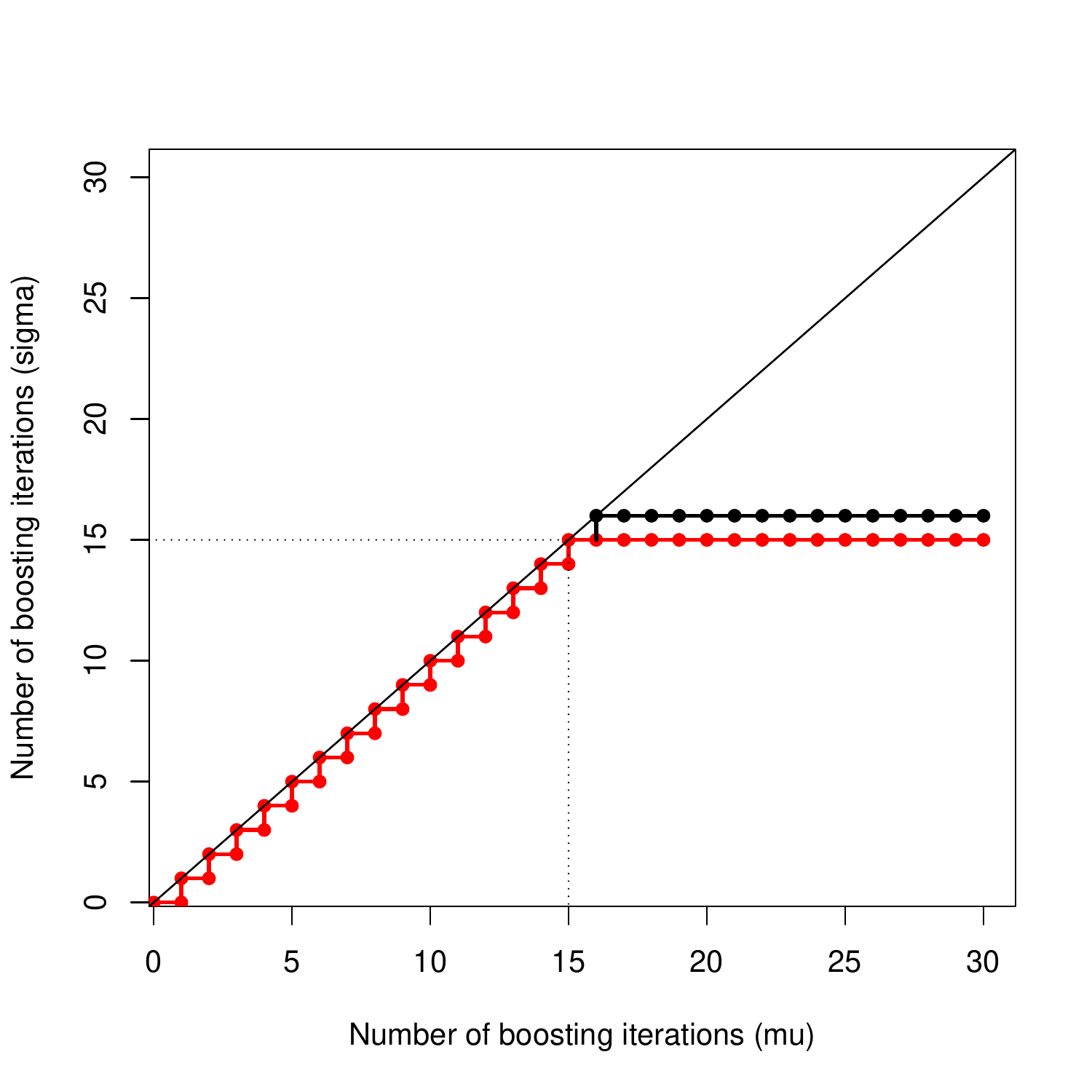}
\caption{\emph{Left:} Comparison between sparse grid (red) and dense $\mu$ grid
  (black horizontal lines in addition to the sparse red grid). For a given
  $m_{\text{stop},\sigma}$, all $m_{\text{stop},\mu}$ values $\geq
  m_{\text{stop},\sigma}$ (i.e., below the bisecting line) can be computed
  without additional computing time. \newline \emph{Right:} Example of the path
  of the iteration count for a model with \code{mstop = c(mu = 30, sigma = 15)}.
  All combinations on the path (red dots) are computed. Until the point where
  $m_{\text{stop}, \mu} = m_{\text{stop}, \sigma}$, we move along the bisecting
  line. Then we stop increasing $m_{\text{stop}, \sigma}$ and increase
  $m_{\text{stop}, \mu}$ only, i.e., we start moving on a horizontal line. Thus,
  all iterations on this horizontal line are computed anyway. Note that it is
  quite expensive to move from the computed model to one with \code{mstop = c(mu
    = 30, sigma = 16)}. One cannot simply increase $m_{\text{stop},\sigma}$ by 1
  but needs to go along the black dotted path.}\label{fig:grid}
\end{figure}

A comparison and an illustration of the sparse and the dense grids can be found
in Figure~\ref{fig:grid} (left). As the additional grid points do not increase
the run time (or only marginally; for an explanation see Figure~\ref{fig:grid},
right), we recommend to always use this option, which is also the default.

\setkeys{Gin}{width = 0.45\textwidth}

The \code{dense\_mu\_grid} option also works for asymmetric grids (e.g.,
\code{make.grid(max = c(mu = 100, sigma = 200))}) and for more than two
parameters (e.g., \code{make.grid(max = c(mu = 100, sigma = 200, nu = 20))}).
For an example in the latter case see the manual of \code{make.grid()}.

Now, we use the dense grid for cross-validation (or subsampling to be more
precise). The computation of the cross-validated risk using \code{cvrisk()} takes
more than one hour on a 64-bit Ubuntu machine using 2 cores.\footnote{By using
  more computing cores or a larger computer cluster the speed can be easily
  increased. The usage of \code{cvrisk()} is practically identical to that of
  \code{cvrisk()} from package \pkg{mboost}. See \citet{Hofner:mboost:2014} for
  details on parallelization and grid computing.} Thus, we only run the
following code if the result does not exist yet.

\begin{Schunk}
\begin{Sinput}
R> ## use multiple cores on non-windows systems:
R> cores <- ifelse(grepl("linux|apple", R.Version()$platform), 2, 1)
R> if (!file.exists("cvrisk/cvr_india.Rda")) {
+      set.seed(1907)    ## set seed for reproducibility
+      folds <- cv(model.weights(mod), type = "subsampling")
+      cvr <- cvrisk(mod, grid = densegrid, folds = folds, mc.cores = cores)
+      save("cvr", file = "cvrisk/cvr_india.Rda")
+  }
\end{Sinput}
\end{Schunk}

We then load the pre-computed results of the cross-validated risk:

\begin{Schunk}
\begin{Sinput}
R> load("cvrisk/cvr_india.Rda")
\end{Sinput}
\end{Schunk}
\vspace{-2em} \hfill $\vardiamond$

\subsection{Methods to extract and display results}\label{sec:methods}

In order to work with the results, methods to extract information both from
boosting models and the corresponding cross-validation results have been
implemented. Fitted \pkg{gamboostLSS} models (i.e., objects of type
\code{"mboostLSS"}) are lists of \code{"mboost"} objects. The most important
distinction from the methods implemented in \pkg{mboost} is the widespread
occurrence of the additional argument \code{parameter}, which enables the user
to apply the function on all parameters of a fitted GAMLSS model or only on one
(or more) specific parameters.

Most importantly, one can extract the coefficients of a fitted model
(\code{coef()}) or plot the effects (\code{plot()}). Different versions of both
functions are available for linear GAMLSS models (i.e., models of class
\code{"glmboostLSS"}) and for non-linear GAMLSS models (e.g., models
with P-splines). Additionally, the user can extract the risk for all iterations
using the function \code{risk()}. Selected base-learners can be extracted using
\code{selected()}. Fitted values and predictions can be obtained by \code{fitted()}
and \code{predict()}. For details and usage examples, see the corresponding
manuals and \citet{Hofner:mboost:2014}. Furthermore, a special function for
marginal prediction intervals is available (\code{predint()}) together with a
dedicated plot function (\code{plot.predint()}).

For cross-validation results (objects of class \code{"cvriskLSS"}), there exists
a function to extract the estimated optimal number of boosting iteration
(\code{mstop()}). The results can also be plotted using a special \code{plot()}
function. Hence, convergence and overfitting behavior can be visually inspected.

In order to increase or reduce the number of boosting steps to the appropriate
number (as e.g., obtained by cross-validation techniques) one can use the subset
operator. If we want to reduce our model, for example, to 10 boosting steps for
the \code{mu} parameter and 20 steps for the \code{sigma} parameter we can use
\begin{Sinput}
R>  mod[c(10, 20)]
\end{Sinput}
This strategy directly alters the object \code{mod}. Thus, no assignment is
needed. This reduces the memory footprint as the object is not duplicated and
works similarly as documented for \pkg{mboost}
\citep[see][]{Hofner:mboost:2014}. However, note that in contrast to
\code{mboost} boosting steps get typically lost when the number of steps is
reduced. Instead of specifying a vector with separate values for each sub-family
one can also use a single value, which then is used for each sub-family (see
Section~\ref{sec:model-fitting}).

\paragraph{Case study (ctd.): Childhood malnutrition in India}

\begin{figure}[h!]
  \centering
\includegraphics{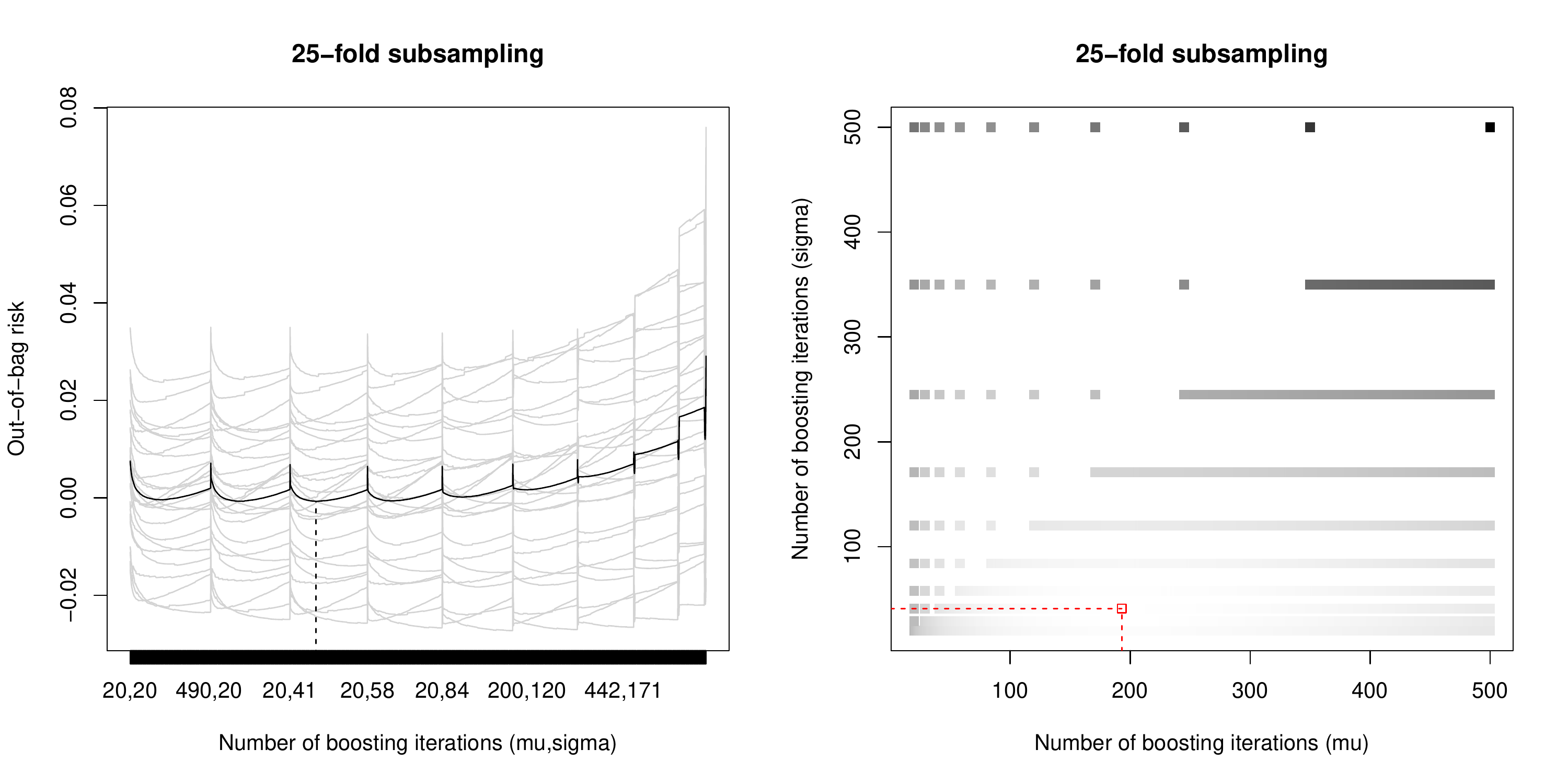}
\caption{Cross-validated risk. The optimal combination of stopping iterations
  (dashed lines) is not at the boundary of the grid. In that case one should
  re-run the cross-validation procedure with different \code{max} values for the
  grid and/or more grid points.}\label{fig:cvr}
\end{figure}

We first inspect the cross-validation results (see Figure~\ref{fig:cvr}):

\begin{Schunk}
\begin{Sinput}
R> par(mfrow = c(1, 2))
R> plot(cvr, type = "lines")
R> plot(cvr, type = "heatmap")
\end{Sinput}
\end{Schunk}

\setkeys{Gin}{width = 0.9\textwidth}

To extract the optimal stopping iteration one can now use
\begin{Schunk}
\begin{Sinput}
R> mstop(cvr)
\end{Sinput}
\begin{Soutput}
   mu sigma 
  193    41 
\end{Soutput}
\end{Schunk}

To use the optimal model, i.e., the model with the iteration number from the
cross-validation, we subset the model to these values by using the subset
operator. Note again that the object \code{mod} is directly altered without assignment.

\begin{Schunk}
\begin{Sinput}
R> mod[mstop(cvr)]
\end{Sinput}
\end{Schunk}

In the next step, the \code{plot()} function can be used to plot the partial
effects. A partial effect is the effect of a certain predictor only, i.e., all
other model components are ignored for the plot. Thus, the reference level of
the plot is arbitrary and even the actual size of the effect might not be
interpretable; only changes and hence the functional form are meaningful. If no
further arguments are specified, all \emph{selected} base-learners are plotted:

\begin{Schunk}
\begin{Sinput}
R> par(mfrow = c(2, 5))
R> plot(mod)
\end{Sinput}
\end{Schunk}

Special base-learners can be plotted using the argument
\code{which}\footnote{Partial matching is used, i.e., one can specify a
  sub-string of the base-learners' names and all matching base-learners are
  selected. Alternatively, one can specify an integer which indicates the number
  of the effect in the model formula.} (to specify the base-learner) and the
argument \code{parameter} (to specify the parameter, e.g., \code{"mu"}). Thus

\begin{figure}[h!]
  \centering
\includegraphics{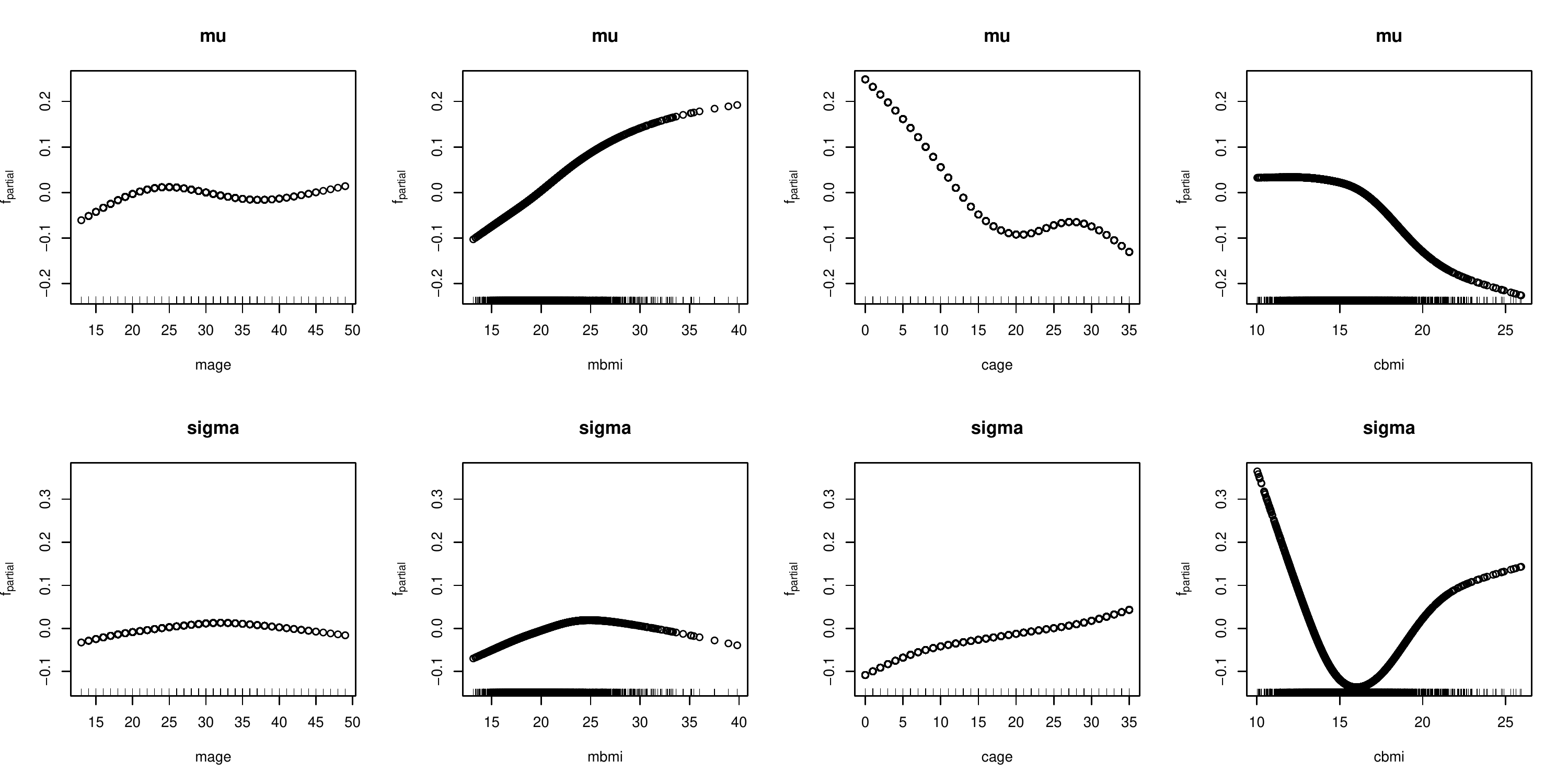}
\caption{Smooth, partial effects of the estimated model with the rescaled
  outcome. The effects for \code{sigma} are estimated and plotted on the
  log-scale (see Equation~\ref{Gaussian:sigma}), i.e., we plot the predictor
  against $\log(\hat{\sigma})$.}\label{fig:smooth_effects}
\end{figure}

\begin{Schunk}
\begin{Sinput}
R> par(mfrow = c(2, 4))
R> plot(mod, which = "bbs")
\end{Sinput}
\end{Schunk}

plots \emph{all} P-spline base-learners irrespective if they where selected or
not (cf.~Figure~\ref{fig:smooth_effects}). The partial effects can be interpreted as
follows:

The age of the mother seems to have a minor impact on stunting for both
the mean effect and the effect on the standard deviation. With increasing BMI of
the mother, the stunting score increases, i.e., the child is better nourished.
At the same time the variability increases until a BMI of roughly 25 and then
decreases again. The age of the child has a negative effect until the age of
approximately 1.5 years (18 months). The variability increases over the complete
range of age. The BMI of the child has a negative effect on stunting, with
lowest variability for an BMI of approximately 16. While all other effects can
be interpreted quite easily, this effect is more difficult to interpret.
Usually, one would expect that a child that suffers from malnutrition also has a
small BMI. However, the height of the child enters the calculation of the BMI in
the denominator, which means that a lower stunting score (i.e., small height)
should lead on average to higher BMI values if the weight of a child is fixed.

\setkeys{Gin}{width = 0.9\textwidth}

If we want to plot the effects of all P-spline base-learners for the $\mu$
parameter, we can use

\begin{Schunk}
\begin{Sinput}
R> plot(mod, which = "bbs", parameter = "mu")
\end{Sinput}
\end{Schunk}

Instead of specifying (sub-)strings for the two arguments one could use integer
values in both cases. For example,

\begin{Schunk}
\begin{Sinput}
R> plot(mod, which = 1:4, parameter = 1)
\end{Sinput}
\end{Schunk}

results in the same plots.

It is also interesting to have a look at marginal prediction intervals using
\code{predint()}. In case of GAMLSS, prediction intervals based on conditional
quantiles can combine the effect of a single predictor variable on various
distribution parameters \citep{mayretal}. For illustration purposes we plot the
influence of the BMI of the child\footnote{Remember that we rescaled the
  outcome. In order to get predictions on the original stunting scale, we need
  to multiply the y-scale by 600.\label{fn:rescaled_outcome}}. To obtain
marginal prediction intervals, we use a grid for the variable of interest, the
mean for all other continuous variables and the modus for categorical variables.

\begin{Schunk}
\begin{Sinput}
R> plot(predint(mod, pi = c(0.8, 0.9), which = "cbmi"),
+       lty = 1:3, lwd = 3, xlab =  "BMI (child)",
+       ylab = "Stunting score", yaxt = "n")
R> ## draw y-axis with a multiplication factor of 600
R> axis(side = 2, at = seq(-1, 1, by = 1/3),
+       labels = seq(-600, 600, by = 200))
\end{Sinput}
\end{Schunk}

\setkeys{Gin}{width = 0.45\textwidth}
\begin{figure}[h!]
  \centering
\includegraphics{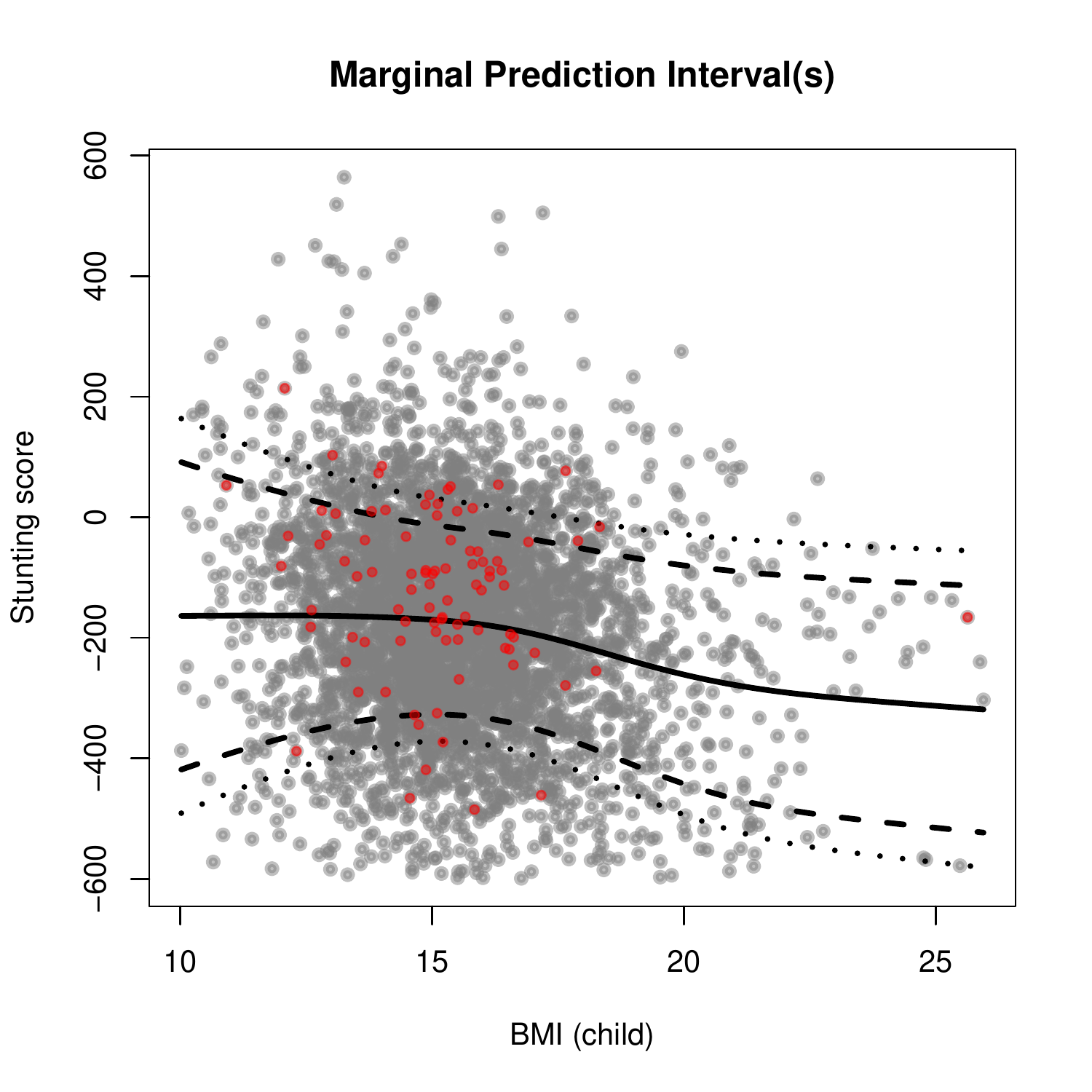}
\caption{80\% (dashed) and 90\% (dotted) marginal prediction intervals for the
  BMI of the children in the district of Greater Mumbai (which is the region
  with the most observations). For all other variables we used average values
  (i.e., a child with average age, and a mother with average age and BMI). The
  solid line corresponds to the median prediction (which equals the mean for
  symmetric distributions such as the Gaussian distribution). Observations from
  Greater Mumbai are highlighted in red.}\label{fig:pred_interval}
\end{figure}

The resulting marginal prediction intervals are displayed in
Figure~\ref{fig:pred_interval}. For the interpretation and evaluation of
prediction intervals, see \cite{MayrPI}.

For the spatial \code{bmrf()} base-learner we need some extra work to plot the
effect(s). We need to obtain the (partial) predicted values per region using
either \code{fitted()} or \code{predict()}\footnote{For \code{bmrf()} base-learners
  one could also plot the coefficients, which constitute the effect estimates
  per region. This isn't true for other bivariate or spatial base-learners such
  as \code{bspatial()} or \code{brad()}.}:
\begin{Schunk}
\begin{Sinput}
R> fitted_mu <- fitted(mod, parameter = "mu", which = "mcdist",
+                      type = "response")
R> fitted_sigma <- fitted(mod, parameter = "sigma", which = "mcdist",
+                         type = "response")
\end{Sinput}
\end{Schunk}

In case of \code{bmrf()} base-learners we then need to aggregate the data for
multiple observations in one region before we can plot the
data\textsuperscript{\ref{fn:rescaled_outcome}}:

\begin{Schunk}
\begin{Sinput}
R> fitted_mu <- tapply(fitted_mu, india$mcdist, FUN = mean) * 600
R> fitted_sigma <- tapply(fitted_sigma, india$mcdist, FUN = mean) * 600
R> plotdata <- data.frame(region = names(fitted_mu),
+                         mean = fitted_mu, sd = fitted_sigma)
R> par(mfrow = c(1, 2))
R> drawmap(data = plotdata, map = india.bnd, regionvar = "region",
+          plotvar = "mean", nrcolors = 19, swapcolors = TRUE,
+          main = "Mean")
R> drawmap(data = plotdata, map = india.bnd, regionvar = "region",
+          plotvar = "sd", nrcolors = 19, swapcolors = TRUE,
+          main = "Standard deviation")
\end{Sinput}
\end{Schunk}

\setkeys{Gin}{width = 0.9\textwidth}
\begin{figure}[h!]
  \centering
\includegraphics{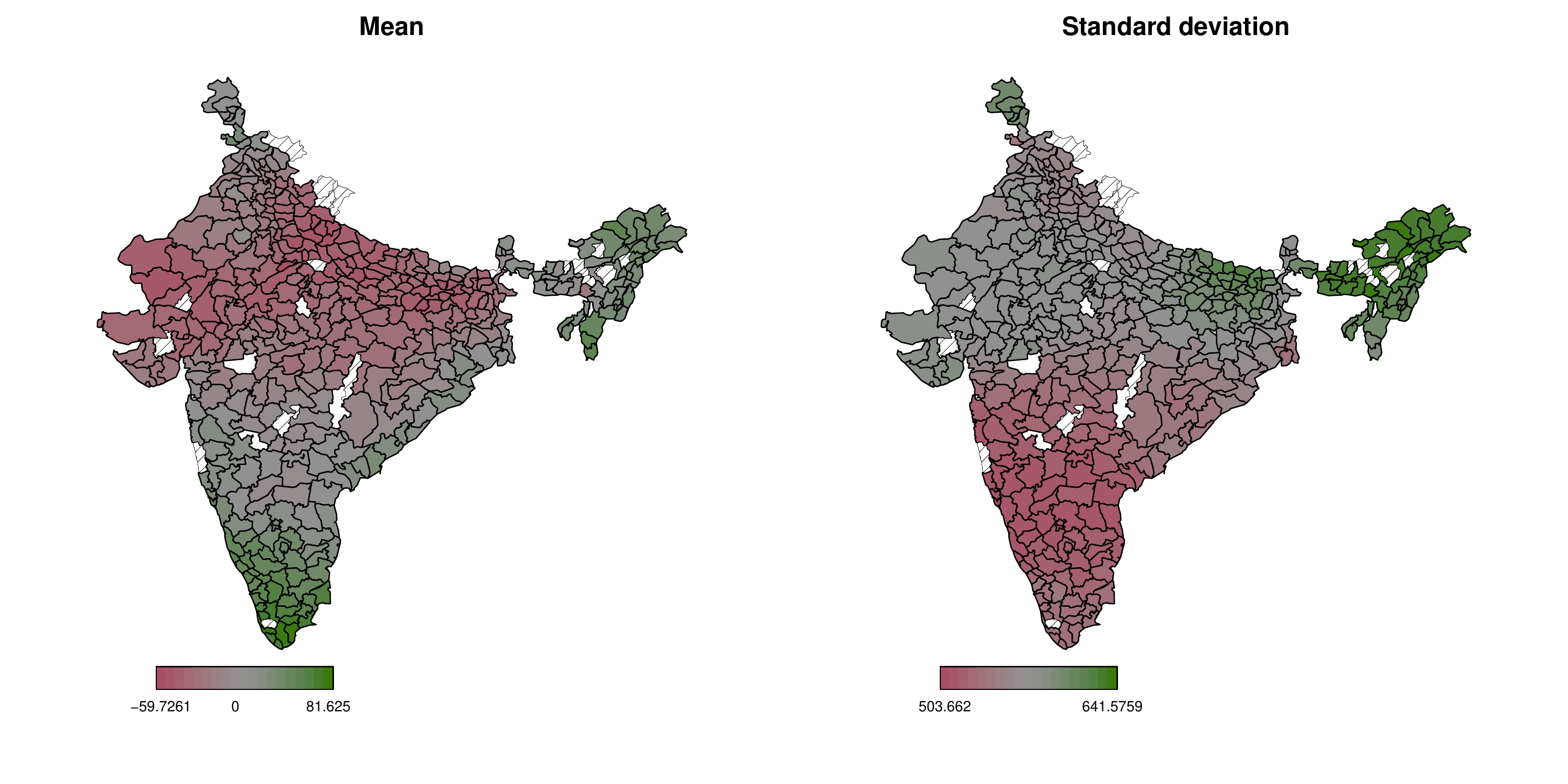}
\caption{Spatial, partial effects of the estimated model. Note that the actual
  effect size is meaningless as we plot partial effects. Only the (size of)
  differences between two regions is meaningful. Dashed regions represent
  regions without data. Note that effect estimates for these regions exist and
  could be extracted. \label{fig:spatial_effects}}
\end{figure}

Figure~\ref{fig:spatial_effects} (left) shows a clear spatial pattern of
stunting. While children in the southern regions like Tamil Nadu and Kerala as
well as in the north-eastern regions around Assam and Arunachal Pradesh seem to
have a smaller risk for stunted growth, the central regions in the north of
India, especially Bihar, Uttar Pradesh and Rajasthan seem to be the most
problematic in terms of stunting due to malnutrition. Since we have also modeled
the scale of the distribution, we can gain much richer information concerning
the regional distribution of stunting: the regions in the south which seem to be
less affected by stunting do also have a lower partial effect with respect to
the expected standard deviation (Figure~\ref{fig:spatial_effects}, right), i.e.,
a reduced standard deviation compared to the average region. This means that not
only the expected stunting score is smaller on average, but that the
distribution in this region is also narrower. This leads to a smaller size of
prediction intervals for children living in that area. In contrast, the regions
around Bihar in the central north, where India shares border with Nepal, do not
only seem to have larger problems with stunted growth but have a positive
partial effect with respect the scale parameter of the conditional distribution
as well. This leads to larger prediction intervals, which could imply a greater
risk for very small values of the stunting score for an individual child in that
region. On the other hand, the larger size of the interval also offers the
chance for higher values and could reflect higher differences between different
parts of the population. \hfill $\vardiamond$

\section{Summary}
\label{sec:summary}

The GAMLSS model class has developed into one of the most flexible tools in
statistical modeling as it can tackle nearly any regression setting of practical
relevance. Boosting algorithms, on the other hand, are one of the most flexible
estimation and prediction tools in the toolbox of a modern statistician.

In this paper, we have presented the \proglang{R} package \pkg{gamboostLSS},
which provides the first implementation of a boosting algorithm for GAMLSS.
Hence, as a combination of boosting and GAMLSS, \pkg{gamboostLSS} combines a
powerful machine learning tool with the world of statistical modeling
\citep{Breiman2001:TwoCultures}, offering the advantage of intrinsic model
choice and variable selection in potentially high-dimensional data situations.
The package also combines the advantages of both \pkg{mboost} (with a
well-established, well-tested modular structure in the back-end) and
\pkg{gamlss} (which implements a large amount of families which are available
via conversion with the \code{as.families()} function).

While the implementation in the \proglang{R} package \pkg{gamlss} (provided by
the inventors of GAMLSS) must be seen as the gold standard for fitting GAMLSS,
the \pkg{gamboostLSS} package offers a flexible alternative, which can be
advantageous, amongst others, in following data settings: (i) models with a
large number of coefficients, where classical estimation approaches become
unfeasible; (ii) data situations where variable selection is of great interest;
(iii) models where a greater flexibility regarding the effect types is needed,
e.g., when spatial, smooth, random, or constrained effects should be included
and selected at the same time.

%%%%%%%%%%%%%%%%%%%%%%%%%%%%%%%%%%%%%%%%%%%%%%%%%%%%%%%%%%%%%%%%%%%%%%%%%%%%%%%%
%% Bibligraphy
%%%%%%%%%%%%%%%%%%%%%%%%%%%%%%%%%%%%%%%%%%%%%%%%%%%%%%%%%%%%%%%%%%%%%%%%%%%%%%%%

\bibliography{literature}

\begin{thebibliography}{39}
\newcommand{\enquote}[1]{``#1''}
\providecommand{\natexlab}[1]{#1}
\providecommand{\url}[1]{\texttt{#1}}
\providecommand{\urlprefix}{URL }
\expandafter\ifx\csname urlstyle\endcsname\relax
  \providecommand{\doi}[1]{doi:\discretionary{}{}{}#1}\else
  \providecommand{\doi}{doi:\discretionary{}{}{}\begingroup
  \urlstyle{rm}\Url}\fi
\providecommand{\eprint}[2][]{\url{#2}}

\bibitem[{Arnold \emph{et~al.}(2009)Arnold, Parasuraman, Arokiasamy, and
  Kothari}]{india_nut}
Arnold F, Parasuraman S, Arokiasamy P, Kothari M (2009).
\newblock \enquote{Nutrition in {I}ndia. {N}ational {F}amily {H}ealth {S}urvey
  ({NFHS-3}), {I}ndia, 2005-06.}
\newblock \emph{Technical report}, Mumbai: International Institute for
  Population Sciences, Calverton, Maryland, USA: ICF Macro.

\bibitem[{Borghi \emph{et~al.}(2006)Borghi, de~Onis, Garza, Van~den Broeck,
  Frongillo, Grummer-Strawn, Van~Buuren, Pan, Molinari, Martorell, Onyango, and
  Martines}]{borghi_who}
Borghi E, de~Onis M, Garza C, Van~den Broeck J, Frongillo EA, Grummer-Strawn L,
  Van~Buuren S, Pan H, Molinari L, Martorell R, Onyango AW, Martines JC (2006).
\newblock \enquote{Construction of the World Health Organization Child Growth
  Standards: Selection of Methods for Attained Growth Curves.}
\newblock \emph{Statistics in Medicine}, \textbf{25}(2), 247--265.

\bibitem[{Breiman(2001)}]{Breiman2001:TwoCultures}
Breiman L (2001).
\newblock \enquote{Statistical Modeling: The Two Cultures (with Discussion).}
\newblock \emph{Statistical Science}, \textbf{16}, 199--231.

\bibitem[{B{\"u}hlmann and Hothorn(2007)}]{BuhlmannHothorn06}
B{\"u}hlmann P, Hothorn T (2007).
\newblock \enquote{Boosting Algorithms: Regularization, Prediction and Model
  Fitting (with Discussion).}
\newblock \emph{Statistical Science}, \textbf{22}, 477--522.

\bibitem[{B{\"u}hlmann and Yu(2003)}]{BuehlmannYu2003}
B{\"u}hlmann P, Yu B (2003).
\newblock \enquote{Boosting with the {$L_2$} Loss: Regression and
  Classification.}
\newblock \emph{Journal of the American Statistical Association},
  \textbf{98}(462), 324--338.

\bibitem[{de~Onis(2006)}]{onis2006child}
de~Onis M (2006).
\newblock \enquote{WHO Child Growth Standards Based on Length/Height, Weight
  and Age.}
\newblock \emph{Acta Paediatrica}, \textbf{95}(S450), 76--85.

\bibitem[{de~Onis \emph{et~al.}(1993)de~Onis, Monteiro, Akre, and
  Clugston}]{who}
de~Onis M, Monteiro C, Akre J, Clugston G (1993).
\newblock \enquote{The Worldwide Magnitude of Protein-Energy Malnutrition: An
  Overview from the {WHO} Global Database on Child Growth.}
\newblock \emph{Bulletin of the World Health Organizationy}, \textbf{71}(6),
  703--712.

\bibitem[{Eilers and Marx(1996)}]{Eilers1996}
Eilers PHC, Marx BD (1996).
\newblock \enquote{Flexible Smoothing with {B}-splines and Penalties (with
  Discussion).}
\newblock \emph{Statistical Science}, \textbf{11}, 89--121.

\bibitem[{Fahrmeir and Kneib(2011)}]{FahrmeirKneib:2011}
Fahrmeir L, Kneib T (2011).
\newblock \emph{Bayesian Smoothing and Regression for Longitudinal, Spatial and
  Event History Data}.
\newblock Oxford University Press.

\bibitem[{Fenske \emph{et~al.}(2013)Fenske, Burns, Hothorn, and
  Rehfuess}]{fenske2013plos}
Fenske N, Burns J, Hothorn T, Rehfuess EA (2013).
\newblock \enquote{Understanding Child Stunting in India: A Comprehensive
  Analysis of Socio-Economic, Nutritional and Environmental Determinants Using
  Additive Quantile Regression.}
\newblock \emph{PloS ONE}, \textbf{8}(11), e78692.

\bibitem[{Fenske \emph{et~al.}(2011)Fenske, Kneib, and
  Hothorn}]{Fenske:2011:JASA}
Fenske N, Kneib T, Hothorn T (2011).
\newblock \enquote{Identifying Risk Factors for Severe Childhood Malnutrition
  by Boosting Additive Quantile Regression.}
\newblock \emph{Journal of the American Statistical Association}, \textbf{106},
  494--510.

\bibitem[{Hastie and Tibshirani(1990)}]{hastietib}
Hastie T, Tibshirani R (1990).
\newblock \emph{Generalized Additive Models}.
\newblock Chapman \& Hall, London.

\bibitem[{Hofner(2011)}]{Hofner:Dissertation:2011}
Hofner B (2011).
\newblock \emph{Boosting in Structured Additive Models}.
\newblock Ph.D. thesis, LMU München.
\newblock Verlag Dr. Hut, München,
  \urlprefix\url{http://nbn-resolving.de/urn:nbn:de:bvb:19-138053}.

\bibitem[{Hofner \emph{et~al.}(2014{\natexlab{a}})Hofner, Kneib, and
  Hothorn}]{Hofner:constrained:2014}
Hofner B, Kneib T, Hothorn T (2014{\natexlab{a}}).
\newblock \enquote{A Unified Framework of Constrained Regression.}
\newblock ArXiv:1403.7118, \urlprefix\url{http://arxiv.org/abs/1403.7118}.

\bibitem[{Hofner \emph{et~al.}(2014{\natexlab{b}})Hofner, Mayr, Fenske, and
  Schmid}]{pkg:gamboostLSS:1.1-0}
Hofner B, Mayr A, Fenske N, Schmid M (2014{\natexlab{b}}).
\newblock \emph{\pkg{gamboostLSS}: Boosting Methods for GAMLSS Models}.
\newblock \proglang{R} package version 1.1-2,
  \urlprefix\url{http://CRAN.R-project.org/package=gamboostLSS}.

\bibitem[{Hofner \emph{et~al.}(2014{\natexlab{c}})Hofner, Mayr, Robinzonov, and
  Schmid}]{Hofner:mboost:2014}
Hofner B, Mayr A, Robinzonov N, Schmid M (2014{\natexlab{c}}).
\newblock \enquote{Model-Based Boosting in \proglang{R} -- {A} Hands-on
  Tutorial Using the \proglang{R} Package \pkg{mboost}.}
\newblock \emph{Computational Statistics}, \textbf{29}, 3--35.

\bibitem[{Hofner \emph{et~al.}(2011)Hofner, M\"{u}ller, and
  Hothorn}]{Hofner:monotonic:2011}
Hofner B, M\"{u}ller J, Hothorn T (2011).
\newblock \enquote{Monotonicity-Constrained Species Distribution Models.}
\newblock \emph{Ecology}, \textbf{92}, 1895--1901.

\bibitem[{Hothorn \emph{et~al.}(2014)Hothorn, B{\"u}hlmann, Kneib, Schmid, and
  Hofner}]{pkg:mboost:2.3-0}
Hothorn T, B{\"u}hlmann P, Kneib T, Schmid M, Hofner B (2014).
\newblock \emph{\pkg{mboost}: Model-Based Boosting}.
\newblock {R} package version 2.3-0,
  \urlprefix\url{http://CRAN.R-project.org/package=mboost}.

\bibitem[{Khondoker \emph{et~al.}(2009)Khondoker, Glasbey, and
  Worton}]{khondoker2009}
Khondoker M, Glasbey C, Worton B (2009).
\newblock \enquote{A Comparison of Parametric and Nonparametric Methods for
  Normalising {cDNA} Microarray Data.}
\newblock \emph{Biometrical Journal}, \textbf{49}(6), 815--823.

\bibitem[{Klein and Moeschberger(2003)}]{klein03}
Klein JP, Moeschberger ML (2003).
\newblock \emph{Survival Analysis: Techniques for Censored and Truncated Data}.
\newblock Second edition. Springer.

\bibitem[{Kneib \emph{et~al.}(2014)Kneib, Heinzl, Brezger, {Sabanes Bove}, and
  Klein}]{pkg:BayesX:0.2-8}
Kneib T, Heinzl F, Brezger A, {Sabanes Bove} D, Klein N (2014).
\newblock \emph{\pkg{BayesX}: \proglang{R} Utilities Accompanying the Software
  Package \pkg{BayesX}}.
\newblock \proglang{R} package version 0.2-8,
  \urlprefix\url{http://CRAN.R-project.org/package=BayesX}.

\bibitem[{Kneib \emph{et~al.}(2009)Kneib, Hothorn, and Tutz}]{kneibetal}
Kneib T, Hothorn T, Tutz G (2009).
\newblock \enquote{Variable Selection and Model Choice in Geoadditive
  Regression Models.}
\newblock \emph{Biometrics}, \textbf{65}, 626--634.

\bibitem[{Kumar \emph{et~al.}(2013)Kumar, Jeyaseelan, Sebastian, Regi, Mathew,
  and Jose}]{bmc_growth}
Kumar V, Jeyaseelan L, Sebastian T, Regi A, Mathew J, Jose R (2013).
\newblock \enquote{New Birth Weight Reference Standards Customised to Birth
  Order and Sex of Babies from South India.}
\newblock \emph{BMC Pregnancy and Childbirth}, \textbf{13}(1), 1--8.

\bibitem[{Mayr \emph{et~al.}(2014)Mayr, Binder, Gefeller, and
  Schmid}]{mayr_boosting_part1}
Mayr A, Binder H, Gefeller O, Schmid M (2014).
\newblock \enquote{The Evolution of Boosting Algorithms -- From Machine
  Learning to Statistical Modelling.}
\newblock \emph{Methods of Information in Medicine}.
\newblock Accepted, \urlprefix\url{http://arxiv.org/abs/1403.1452}.

\bibitem[{Mayr \emph{et~al.}(2012{\natexlab{a}})Mayr, Fenske, Hofner, Kneib,
  and Schmid}]{mayretal}
Mayr A, Fenske N, Hofner B, Kneib T, Schmid M (2012{\natexlab{a}}).
\newblock \enquote{Generalized Additive Models for Location, Scale and Shape
  for High Dimensional Data -- A Flexible Approach Based on Boosting.}
\newblock \emph{Journal of the Royal Statistical Society, Series C},
  \textbf{61}, 403--427.

\bibitem[{Mayr \emph{et~al.}(2012{\natexlab{b}})Mayr, Hofner, and
  Schmid}]{Mayr:mstop:2012}
Mayr A, Hofner B, Schmid M (2012{\natexlab{b}}).
\newblock \enquote{The Importance of Knowing when to Stop -- A Sequential
  Stopping Rule for Component-Wise Gradient Boosting.}
\newblock \emph{Methods of Information in Medicine}, \textbf{51}, 178--186.

\bibitem[{Mayr \emph{et~al.}(2012{\natexlab{c}})Mayr, Hothorn, and
  Fenske}]{MayrPI}
Mayr A, Hothorn T, Fenske N (2012{\natexlab{c}}).
\newblock \enquote{Prediction Intervals for Future {BMI} Values of Individual
  Children -- A Non-Parametric Approach by Quantile Boosting.}
\newblock \emph{BMC Medical Research Methodology}, \textbf{12}(6).

\bibitem[{{\proglang{R} Core Team}(2014)}]{R:3.1.0}
{\proglang{R} Core Team} (2014).
\newblock \emph{\proglang{R}: A Language and Environment for Statistical
  Computing}.
\newblock \proglang{R} Foundation for Statistical Computing, Vienna, Austria.
\newblock {ISBN} 3-900051-07-0. Software version 3.1.0,
  \urlprefix\url{http://www.R-project.org/}.

\bibitem[{Rigby and Stasinopoulos(2005)}]{rs}
Rigby RA, Stasinopoulos DM (2005).
\newblock \enquote{Generalized Additive Models for Location, Scale and Shape
  (with Discussion).}
\newblock \emph{Applied Statistics}, \textbf{54}, 507--554.

\bibitem[{Rigby and Stasinopoulos(2013)}]{rigby:smoothing:2013}
Rigby RA, Stasinopoulos DM (2013).
\newblock \enquote{Automatic Smoothing Parameter Selection in {GAMLSS} with an
  Application to Centile Estimation.}
\newblock \emph{Statistical Methods in Medical Research}.
\newblock Available Online.

\bibitem[{Schmid \emph{et~al.}(2010)Schmid, Potapov, Pfahlberg, and
  Hothorn}]{schmidetal}
Schmid M, Potapov S, Pfahlberg A, Hothorn T (2010).
\newblock \enquote{Estimation and Regularization Techniques for Regression
  Models with Multidimensional Prediction Functions.}
\newblock \emph{Statistics and Computing}, \textbf{20}(2), 139--150.

\bibitem[{Schmid \emph{et~al.}(2013)Schmid, Wickler, Maloney, Mitchell, Fenske,
  and Mayr}]{schmid2013beta}
Schmid M, Wickler F, Maloney KO, Mitchell R, Fenske N, Mayr A (2013).
\newblock \enquote{Boosted Beta Regression.}
\newblock \emph{PloS ONE}, \textbf{8}(4), e61623.

\bibitem[{Serinaldi and Kilsby(2012)}]{serinaldi2012}
Serinaldi F, Kilsby CG (2012).
\newblock \enquote{A Modular Class of Multisite Monthly Rainfall Generators for
  Water Resource Management and Impact Studies.}
\newblock \emph{Journal of Hydrology}, \textbf{464--465}, 528--540.

\bibitem[{Sobotka and Kneib(2012)}]{sobotka12}
Sobotka F, Kneib T (2012).
\newblock \enquote{Geoadditive Expectile Regression.}
\newblock \emph{Computational Statistics \& Data Analysis}, \textbf{56},
  755--767.

\bibitem[{Stasinopoulos and Rigby(2007)}]{gamlss:jss:2007}
Stasinopoulos DM, Rigby RA (2007).
\newblock \enquote{Generalized Additive Models for Location Scale and Shape
  ({GAMLSS}) in \proglang{R}.}
\newblock \emph{Journal of Statistical Software}, \textbf{23(7)}.

\bibitem[{Stasinopoulos and Rigby(2014{\natexlab{a}})}]{pkg:gamlss:4.3-0}
Stasinopoulos M, Rigby B (2014{\natexlab{a}}).
\newblock \emph{\pkg{gamlss}: Generalized Additive Models for Location Scale
  and Shape}.
\newblock \proglang{R} package version 4.3-0,
  \urlprefix\url{http://CRAN.R-project.org/package=gamlss}.

\bibitem[{Stasinopoulos and Rigby(2014{\natexlab{b}})}]{pkg:gamlss.dist:4.3-0}
Stasinopoulos M, Rigby B (2014{\natexlab{b}}).
\newblock \emph{\pkg{gamlss.dist}: Distributions to be Used for GAMLSS
  Modelling}.
\newblock \proglang{R} package version 4.3-0,
  \urlprefix\url{http://CRAN.R-project.org/package=gamlss.dist}.

\bibitem[{van Ogtrop \emph{et~al.}(2011)van Ogtrop, Vervoort, Heller,
  Stasinopoulos, and Rigby}]{ogtrop2011}
van Ogtrop FF, Vervoort RW, Heller GZ, Stasinopoulos DM, Rigby RA (2011).
\newblock \enquote{Long-Range Forecasting of Intermittent Streamflow.}
\newblock \emph{Hydrology and Earth System Sciences}, \textbf{8}, 681--713.

\bibitem[{Villarini \emph{et~al.}(2009)Villarini, Smith, Serinaldi, Bales,
  Bates, and Krajewski}]{villarini2009}
Villarini G, Smith J, Serinaldi F, Bales J, Bates P, Krajewski W (2009).
\newblock \enquote{Flood Frequency Analysis for Nonstationary Annual Peak
  Records in an Urban Drainage Basin.}
\newblock \emph{Advances in Water Resources}, \textbf{32}, 1255--1266.

\end{thebibliography}

%%%%%%%%%%%%%%%%%%%%%%%%%%%%%%%%%%%%%%%%%%%%%%%%%%%%%%%%%%%%%%%%%%%%%%%%%%%%%%%%
%% Appendix
%%%%%%%%%%%%%%%%%%%%%%%%%%%%%%%%%%%%%%%%%%%%%%%%%%%%%%%%%%%%%%%%%%%%%%%%%%%%%%%%

\clearpage
\appendix

\section[The gamboostLSS algorithm]{The \emph{gamboostLSS} algorithm}\label{algorithm}

Let $\bm{\theta} = (\theta_k)_{k = 1,\ldots,K}$ be the vector of distribution
parameters of a GAMLSS, where $\theta_k = g_k^{-1}(\eta_{\theta_k})$ with
parameter-specific link functions $g_k$ and additive predictor
$\eta_{\theta_k}$. The \emph{gamboostLSS} algorithm \citep{mayretal} circles
between the different distribution parameters $\theta_k,\, k=1, \ldots, K,$ and
fits all base-learners $h(\cdot)$ separately to the negative partial derivatives
of the loss function, i.e., in the GAMLSS context to the partial derivatives of
the log-likelihood with respect to the additive predictors $\eta_{\theta_k}$,
i.e., $\frac{\partial}{\partial \eta_{\theta_k}} l(\bm{y}, \bm{\theta})$.

\begin{enumerate}
\item[] \textbf{Initialize}
  \begin{enumerate}
  \item[(1)] Set the iteration counter $m := 0$. Initialize the additive
    predictors $\hat{\eta}_{\theta_{k,i}}^{[m]},\, k = 1, \ldots, K,\, i=1,
    \ldots, n,$ with offset values, e.g. $\hat{\eta}_{\theta_{k,i}}^{[0]} \equiv
    \underset{c}{\operatorname{argmax}} \sum_{i=1}^n l(y_i, \theta_{k,i} = c)$.
  \item[(2)] For each distribution parameter $\theta_k$, $k=1,\ldots, K$,
    specify a set of base-learners: i.e., for parameter $\theta_k$ by $h_{k,1}
    (\cdot),\ldots,h_{k,p_k} (\cdot)$, where $p_k$ is the cardinality of the set
    of base-learners specified for $\theta_k$.
  \end{enumerate}
\item[] \textbf{Boosting in multiple dimensions}
  \begin{enumerate}
  \item[(3)] \textbf{Start} a new boosting iteration: increase $m$ by 1 and set
    $k := 0$.
  \item[(4)]
    \begin{enumerate}
    \item[(a)] Increase $k$ by 1. \\
      \textbf{If} $m > m_{\rm{stop},\emph{k}}$ proceed to
      step 4(e).\\
      \textbf{Else} compute the partial derivative $\frac{\partial}{\partial
        \eta_{\theta_k}} l(y,\bm{\theta})$ and plug in the current estimates
      $\hat{\bm{\theta}}_i^{[m-1]} = \left( \hat{\theta}_{1,i}^{[m-1]}, \ldots,
        \hat{\theta}_{K,i}^{[m-1]} \right) = \left(
        g^{-1}_1(\hat{\eta}_{\theta_{1,i}}^{[m-1]}), \ldots,
        g^{-1}_K(\hat{\eta}_{\theta_{K,i}}^{[m-1]}) \right)$:
      \begin{equation*}
        u^{[m-1]}_{k,i} = \left. \frac{\partial}{\partial \eta_{\theta_k}} l(y_i,
          \bm{\theta})\right|_{\bm{\theta} = \hat{\bm{\theta}}_i^{[m-1]}} ,\, i = 1,\ldots, n.
      \end{equation*}
    \item[(b)] \textbf{Fit} each of the base-learners contained in the set of
      base-learners specified for the parameter $\theta_k$ in step (2) to the
      gradient vector $\bm{u}^{[m-1]}_k$.
    \item[(c)] \textbf{Select} the base-learner $j^*$ that best fits the
      partial-derivative vector according to the least-squares criterion, i.e.,
      select the base-learner $h_{k,j^*}$ defined by
      \begin{equation*}
        j^* = \underset{1 \leq j \leq p_k}{\operatorname{argmin}}\sum_{i=1}^n (u_{k,i}^{[m-1]} - h_{k,j}(\cdot))^2 \ .
      \end{equation*}
    \item[(d)] \textbf{Update} the additive predictor $\eta_{\theta_k}$ as
      follows:
      \begin{equation*}
        \hat{\eta}_{\theta_k}^{[m-1]} := \hat{\eta}_{\theta_k}^{[m-1]} + \nu_{\text{sl}} \cdot h_{k,j^*}(\cdot)\ ,
      \end{equation*}
      where $\nu_{\text{sl}}$ is a small step-length ($0 <
        \nu_{\text{sl}} \ll 1$).
    \item[(e)] Set $\hat{\eta}_{\theta_k}^{[m]} :=
      \hat{\eta}_{\theta_k}^{[m-1]}$.
    \item[(f)] \textbf{Iterate} steps 4(a) to 4(e) for $k=2,\ldots, K$.
    \end{enumerate}
  \end{enumerate}

\item[] \textbf{Iterate}
  \begin{enumerate}
  \item[(5)] Iterate steps 3 and 4 until  $m > m_{\text{stop},k}$ for all
    $k=1,\ldots, K$.
  \end{enumerate}
\end{enumerate}

\section{Data pre-processing and stabilization of gradients}\label{sec:stab-ngrad}

As the \emph{gamboostLSS} algorithm updates the parameter estimates in turn by
optimizing the gradients, it is important that these are comparable for
all GAMLSS parameters. Consider for example the standard Gaussian distribution
where the gradients of the log-likelihood with respect to $\eta_{\mu}$ and
$\eta_{\sigma}$ are
\begin{equation*}
   \frac{\partial }{\partial \eta_{\mu}}\, l(y_i, g_{\mu}^{-1}(\eta_{\mu}), \hat{\sigma}) =
  \frac{ y_i - \eta_{\mu i}}{\hat{\sigma}_i^2},
\end{equation*}
with identity link, i.e., $g_{\mu}^{-1}(\eta_{\mu}) = \eta_{\mu}$, and
\begin{equation*}
  \frac{\partial }{\partial \eta_{\sigma}}\, l(y_i, \hat{\mu}, g_{\sigma}^{-1}(\eta_{\sigma})) =
  -1 + \frac{(y_i - \hat{\mu}_i)^2}{\exp(2\eta_{\sigma i})},
\end{equation*}
with log link, i.e., $g_{\sigma}^{-1}(\eta_{\sigma}) = \exp(\eta_{\sigma})$.

For small values of $\hat{\sigma}_i$, the gradient vector for $\mu$ will hence
inevitably become huge, while for large variances it will become very small. As
the base-learners are directly fitted to this gradient vector, this will have a
dramatic effect on convergence speed. Due to imbalances regarding the range of
$\frac{\partial }{\partial \eta_{\mu}} l(y_i, \mu, \sigma)$ and $\frac{\partial
}{\partial \eta_{\sigma}} l(y_i, \mu, \sigma)$, a potential bias might be
induced when the algorithm becomes so unstable that it does not converge to the
optimal solution (or converges very slowly).

Another but related issue might be the different ranges of the gradient and the
outcome itself. In our example, the range of the outcome variable
\code{stunting} is approximately $\pm 600$, while the range of the gradient is
approximately $\pm 2$. As a consequence it takes very long to adapt to the
correct data range by fitting the gradient. In this case, standardization of the
outcome (and/or a larger step size $\nu_{\text{sl}}$) might be highly beneficial
for a faster convergence.

Consequently, there are two methods to achieve comparable gradients and thus
more stable model estimates and faster convergence. First, the user can try to
standardize the variables, especially the outcome. Second, one can use
standardized gradients, where in \textbf{each step} the gradient is divided by
its median absolute deviation, i.e., it is divided by
\begin{equation}
  \label{eq:MAD}
  \text{MAD} = \text{median}_i(|u_{k,i} - \text{median}_j(u_{k,j})|),
\end{equation}
where $u_{k,i}$ is the  gradient of the $k$th GAMLSS parameter in the
current boosting step $i$. This can be activated by setting the options to
\begin{Sinput}
R> options(gamboostLSS_stab_ngrad = TRUE)
\end{Sinput}
and deactivated by
\begin{Sinput}
R> options(gamboostLSS_stab_ngrad = FALSE)
\end{Sinput}
Currently, the latter is the default. Both methods can be used at the same time
if this is required by the data set. \pagebreak

\section{Additional Families}\label{sec:additional-families}

Table~\ref{tab:gamlss_families} gives an overview of common, additional GAMLSS
distributions and GAMLSS distributions with a different parametrization than in
\pkg{gamboostLSS}. For a comprehensive overview see the distribution tables
available at \url{www.gamlss.org} and the documentation of the
\pkg{gamlss.dist} package \citep{pkg:gamlss.dist:4.3-0}. Note that
\pkg{gamboostLSS} works only for more-parametric distributions, while in
\pkg{gamlss.dist} also a few one-parametric distributions are implemented. In
this case the \code{as.families()} function will construct a corresponding
\code{"boost\_family"} which one can use as \code{family} in \pkg{mboost} (a
corresponding advice is given in a warning message).

\begin{landscape}
  \begin{table}[h!]
    \centering
    \caption{Overview of common, additional GAMLSS distributions that can be used via
      \code{as.families()} in \pkg{gamboostLSS}. For every modeled
      distribution parameter, the corresponding link-function is displayed.} \label{tab:gamlss_families}
    \begin{tabular}{llllllllp{0.5\textwidth}}
      \toprule
      & &  Name & Response & $\mu$ & $\sigma$  & $\nu$  & $\tau$  & Note\\
      \cmidrule{2-9}

      \multicolumn{9}{l}{\textbf{Continuous response}} \\
      & Generalized $t$  & \code{GT} & cont.   & id & log & log& log & \\
      & Box-Cox $t$  & \code{BCT} & cont.   & id & log & id& log& \\
      & Gumbel & \code{GU} & cont.  & id & log & & & For moderately skewed data.\\
      & Reverse Gumbel & \code{RG} & cont.  & id & log & & &  Extreme value distribution.\\
      \cmidrule{2-9}

      \multicolumn{9}{l}{\textbf{Continuous non-negative response} (without censoring)} \\
      & Gamma & \code{GA} & cont. $>0$  & log & log & & & Also implemented as
      \code{GammaLSS()}$\,^{a,b}$.\\
      & Inverse Gamma & \code{IGAMMA} & cont. $>0$  & log & log & & & \\
      & Zero-adjusted Gamma & \code{ZAGA} & cont. $\geq 0$    & log & log& logit
      & & Gamma, additionally allowing for zeros.\\
      & Inverse Gaussian & \code{IG} & cont. $>0$   & log & log & & & \\
      & Log-normal & \code{LOGNO} & cont. $>0$  & log & log & & & For positively skewed data.\\
      & Box-Cox Cole and Green & \code{BCCG} & cont. $>0$  & id & log & id & & For positively and negatively skewed data.\\
      & Pareto & \code{PARETO2} & cont. $>0$  & log & log & & & \\
      & Box-Cox power exponential & \code{BCPE} & cont. $>0$  & id & log & id & log & Recommended for child growth centiles.\\
      \cmidrule{2-9}

      \multicolumn{9}{l}{\textbf{Fractions and bounded continuous response}} \\
      & Beta & \code{BE} & $\in (0,1)$  & logit & logit & & & Also implemented
      as \code{BetaLSS()}$\,^{a,c}$.\\
      & Beta inflated & \code{BEINF} & $\in [0,1]$  & logit & logit & log &log
      & Beta, additionally allowing for zeros and ones.\\
      \cmidrule{2-9}

      \multicolumn{9}{l}{\textbf{Models for count data}} \\
      & Beta binomial & \code{BB} & count  & logit & log &  &  &   \\
      & Negative binomial & \code{NBI} & count  & log & log &  &  & For
      over-dispersed count data; also implemented as
      \code{NBinomialLSS()}$\,^{a,d}$.\\
      \bottomrule
      \multicolumn{9}{p{1.3\textwidth}}{\small%
        $^a$ The parametrizations of the distribution functions in \pkg{gamboostLSS} and \pkg{gamlss.dist} differ with respect to the variance.\newline
          $\,^b$ \code{GammaLSS(mu, sigma)} has $\VAR(y|x) = \text{\code{mu}}^2/\text{\code{sigma}}$,
          and \code{as.families(GA)(mu, sigma)} has $\VAR(y|x) = \text{\code{sigma}}^2\cdot \text{\code{mu}}^2$.\newline
          $\,^c$ \code{BetaLSS(mu, phi)} has $\VAR(y|x) = \text{\code{mu}}\cdot(1- \text{\code{mu}}) \cdot(1 + \text{\code{phi}})^{-1}$,
          and \code{as.families(BE)(mu, sigma)} has $\VAR(y|x) =  \text{\code{mu}}\cdot(1- \text{\code{mu}}) \cdot\text{\code{sigma}}^2$\newline
          $\,^d$ \code{NBinomialLSS(mu, sigma)} has $\VAR(y|x) = \text{\code{mu}}+ 1/\text{\code{sigma}}\cdot\text{\code{mu}}^2$,
          and \code{as.families(NBI)(mu, sigma)} has $\VAR(y|x) = \text{\code{mu}}+\text{\code{sigma}}\cdot\text{\code{mu}}^2$}
    \end{tabular}
  \end{table}
\end{landscape}

\end{document}